\documentclass[a4paper,12pt,britishm,texlive=2016]{article}

\RequirePackage{graphicx}
\RequirePackage{subfig}
\RequirePackage{mhchem}
\RequirePackage{url}
\RequirePackage{xcolor}
\RequirePackage{bbold}
\RequirePackage{placeins}

\newcommand{\TeV}{\ensuremath{\text{Te\kern -0.1em V}}}
\newcommand{\GeV}{\ensuremath{\text{Ge\kern -0.1em V}}}
\newcommand{\MeV}{\ensuremath{\text{Me\kern -0.1em V}}}

\makeatletter
\g@addto@macro\bfseries\boldmath
\makeatother

\title{SR-GAN for SR-gamma:\\ super resolution of photon calorimeter \\ images at collider experiments\\}

\author{Johannes Erdmann$^1$, Aaron van der Graaf$^2$,\\
  Florian Mausolf$^1$, Olaf Nackenhorst$^2$}

\date{\small
  $^1$ RWTH Aachen University, III. Physikalisches Institut A, Aachen, Germany\\
  $^2$ TU Dortmund University, Fakult\"at f\"ur Physik, Dortmund, Germany
  }

\begin{document}

\maketitle

\begin{abstract}
  We study single-image super-resolution algorithms for photons at collider experiments based on generative adversarial networks.
  We treat the energy depositions of simulated electromagnetic showers of photons and neutral-pion decays in a toy electromagnetic calorimeter as 2D images and we train super-resolution networks to generate images with an artificially increased resolution by a factor of four in each dimension.
  The generated images are able to reproduce features of the electromagnetic showers that are not obvious from the images at nominal resolution.
  Using the artificially-enhanced images for the reconstruction of shower-shape variables and of the position of the shower center results in significant improvements. 
  We additionally investigate the utilization of the generated images as a pre-processing step for deep-learning photon-identification algorithms and observe improvements in the case of training samples of small size. 
\end{abstract}

\section{Introduction}
\label{sec:introduction}

The interaction of high-energy particles with matter results in complex signatures in the detectors at particle colliders, such as the LHC~\cite{Evans:2008zzb}.
The reconstruction and identification of particles from the detector signatures are crucial to carry out physics analyses.
An important particle is the photon, which appears for example in the diphoton decay of the Higgs boson at the ATLAS and CMS experiments~\cite{ATLAS:2012yve,CMS:2012qbp}, as a probe of heavy-ion collisions at the ALICE experiment~\cite{ALICE:2015xmh} or as a decay product of rare $B$-meson decays at the LHCb experiment~\cite{LHCb:2021awg}.
The main signature of a high-energy photon is an electromagnetic shower in the calorimeter.

At hadron colliders, a main background source for photons are electromagnetic decays of high-energy mesons, most prominently, the decay $\pi^0\rightarrow\gamma\gamma$, as neutral pions are copiously produced in the fragmentation of quarks and gluons.
The signature of such a high-energy meson decay often produces a ``fake single photon'', because the large Lorentz boost leads to a small average distance between the photons from its decay.
This results in a signature that is very similar to the signature of a real single photon.
Distinguishing real from fake single photons is hence challenging and an important design consideration for electromagnetic calorimeters.
Key to distinguishing these two signatures is a high spatial resolution that is achieved by segmenting the calorimeter along pseudorapidity\footnote{The pseudorapidity is defined as $\eta = -\ln\left(\tan\left(\theta/2\right)\right)$, where $\theta$ is the polar angle.} $\eta$ and azimuthal angle $\phi$.

In this work, we study how single-image super resolution (SR)~\cite{SR_survey} based on deep neural networks~\cite{Yang_2019} can help in the reconstruction of photon and $\pi^0$ signatures. 
Such deep-learning algorithms were pioneered~\cite{SISRDL} in the field of image processing and further developed~\cite{SISRGAN} using the concept of generative adversarial networks (GAN)~\cite{DBLP:journals/corr/GoodfellowPMXWOCB14}.
They aim at learning an SR version of a low-resolution (LR) image based on its high-resolution (HR) counterpart, where the number of pixels is identical for the SR and HR images.
We use a neural network inspired by the Enhanced Super-Resolution Generative Adversarial Networks (ESRGAN)~\cite{DBLP:journals/corr/abs-1809-00219}.
While the generator of the GAN produces artificial SR images from input LR images, the discriminator of the GAN aims to distinguish SR and HR images.
By combining the generator and discriminator loss into a common loss term, the generated SR images are expected to become more and more realistic during the GAN training.

We treat the calorimeter signatures of photons and neutral pions as the LR images, i.e.~the LR images correspond to the granularity of an actual calorimeter.
We use simulations of LR images and their corresponding HR counterparts, which have a finer calorimeter segmentation, to train the ESRGAN.
Previous applications of super resolution in the field of particle physics focussed on energy and directional reconstruction of charged and neutral pions~\cite{DiBello:2020bas}, on the reconstruction of jet substructure~\cite{Baldi:2020hjm}, and recently, on refining fast calorimeter simulations~\cite{pang2023supercalo}.
We focus on the particularly relevant use case of photon identification and reconstruction.
We use a toy calorimeter inspired by the electromagnetic calorimeter of the CMS detector~\cite{CMS:2008xjf} with a realistic simulation of the particle interaction with matter using {\sc Geant4}~\cite{GEANT4:2002zbu}.
We study whether the generated SR images provide advantages compared to only using their LR counterparts for benchmark applications in photon--neutral-pion separation and in the directional reconstruction of the photons.
The latter application is especially important for the reconstruction of invariant masses from photon signatures, such as in $H\rightarrow\gamma\gamma$.
We comment on useful strategies for a stable GAN training and on how the additional physics information from the HR images may help in stabilizing photon classifier trainings in case of limited number of training samples.

\section{Simulated samples}
\label{sec:simulation}

\begin{figure}[t]
    \centering
    \includegraphics[width=0.5\textwidth]{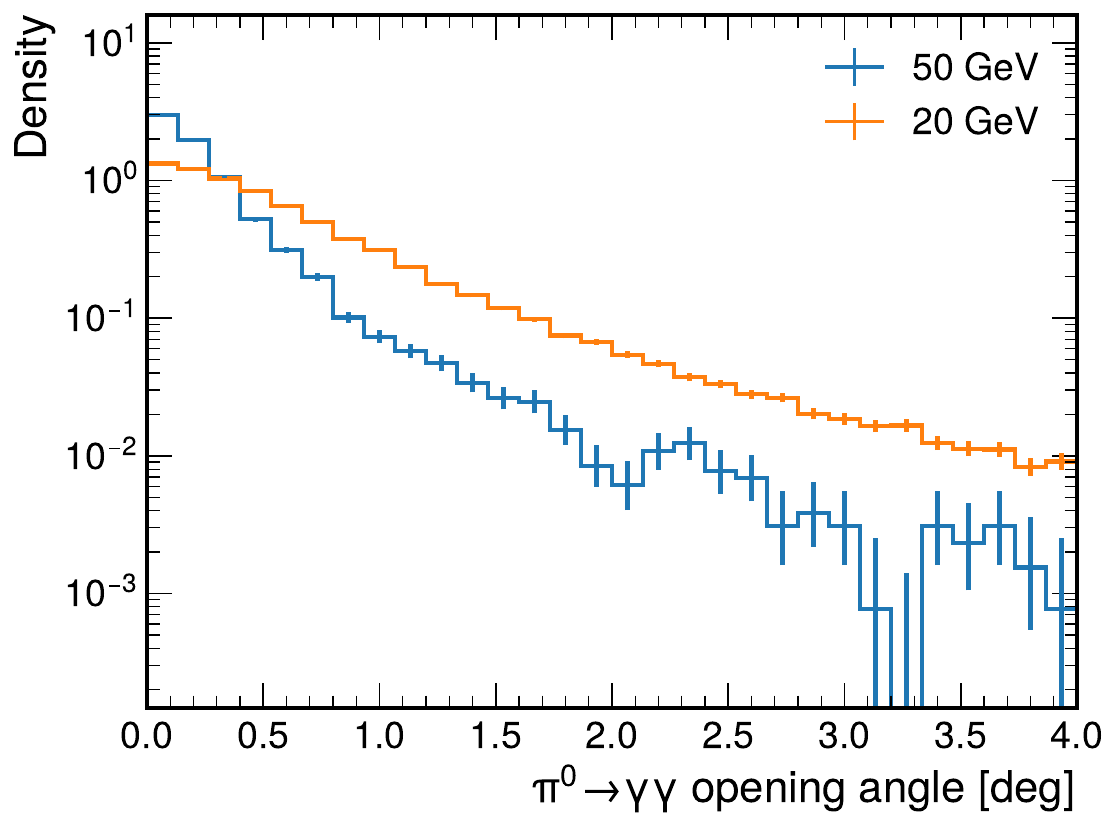}
    \caption{Normalized distributions of the angle between the photons of the $\pi^0 \to\gamma\gamma$ decays at $20\,\GeV$ and $50\,\GeV$ in the lab frame.
    The overflow is not included.
    For both energies, the majority of pions decay into photons that are closer to each other than $1\,\deg$, which corresponds to a separation at the calorimeter front of less than one crystal width in the LR case.
    }
    \label{fig:simulation/opening_angles}
\end{figure}

We simulate a toy calorimeter that is inspired by the electromagnetic barrel calorimeter of the CMS detector.
We use the framework of the CaloGAN paper~\cite{Paganini:2017dwg} based on {\sc Geant4}~\texttt{10.6.2} to simulate PbWO$_4$ scintillating crystals with a length of $230\,\text{mm}$ and a front face of $22\times22\,\text{mm}^2$.
The front of the calorimeter is placed at a distance of $1.29\,\text{m}$ from a {\sc Geant4} particle gun.
The particle gun produces mono-energetic photons and neutral pions with their direction perpendicular to the calorimeter front face.
In order to avoid that all particles are directed at the exact center of the calorimeter, the position of the source is smeared in the plane parallel to the calorimeter front using a Gaussian distribution of width $44\,\text{mm}$, which corresponds to the size of two crystals.
Two different energies of $20\,\GeV$ and $50\,\GeV$ are simulated, which are chosen to be at the lower end of reconstructable photon energies at the LHC and of the order of typical photon energies from Higgs-boson decays.
The $\pi^0$ mesons decay into a pair of photons with an angular separation between them as shown in Fig.~\ref{fig:simulation/opening_angles}. 
In both setups, the majority of pion decays produces photons closer to each other than $1\,\deg$, which results in a separation at the calorimeter front of less than one crystal width.
Due to the larger Lorentz boost, the decays at an energy of $50\,\GeV$ are more collimated on average than in the $20\,\GeV$ case.
We remove simulated pions where the angle between the photons exceeds $2\,\deg$, because their decays often lead to two well-separated photons even in the LR case.
This angular selection retains around $94\,\%$ and $99\,\%$ of the simulated pions with an energy of $20\,\GeV$ and $50\,\GeV$, respectively.
We did not simulate the calorimeter noise, a magnetic field or material upstream of the calorimeter.

The LR images consist of a grid of $24\times 24$ crystals.
The HR images have a segmentation that is $4\times 4$ finer, i.e.~$96\times 96$ smaller crystals. 
In order to maintain a one-to-one correspondence between LR and HR images, only HR images are simulated.
The LR images are then derived by down-sampling the HR images, wherein the energy sum of each $4\times4$ HR patch is assigned as the corresponding LR crystal's energy.

\begin{figure}[p]
    \centering
    \includegraphics[width=0.9\textwidth]{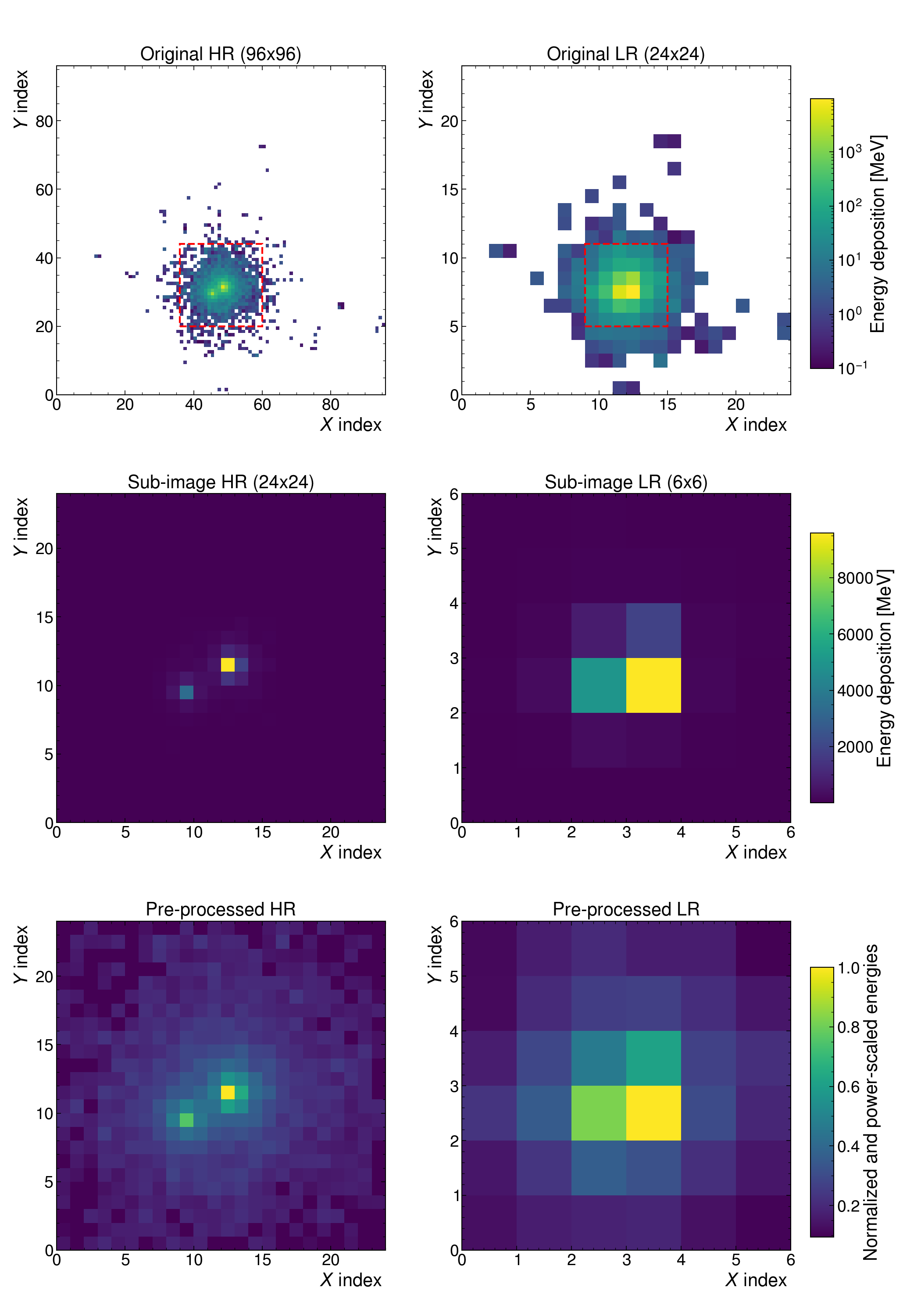}
    \caption{Visualization of the pre-processing of the calorimeter images, shown for a $20\,\GeV$ pion example. 
    In the first row, the simulated image is shown in HR (left) and in LR (right) with a logarithmic colorbar.
    The selected sub-images are marked in red and displayed in the second row with a linear colorbar.
    The third row shows the normalized and power-scaled images with a linear colorbar.}
    \label{fig:simulation/Preprocessing}
\end{figure}

Before being passed to the networks, the calorimeter images are pre-processed. 
The two pre-processing steps are visualized in Fig.~\ref{fig:simulation/Preprocessing} for a HR pion image and its corresponding LR counterpart.
In a first step (going from the first to the second row in the figure), the size of the images is reduced in order to decrease the computational complexity of the super-resolution networks.
The width of $2.2\,\text{cm}$ of the LR calorimeter crystals corresponds to approximately one Moli\`{e}re radius in PbWO$_4$, causing photons to deposit most of their energy within a small number of crystals.
Therefore, we select the $6\times6$ sub-image that contains the largest sum of energy within our LR simulation of $24\times24$ crystals.
For the HR images, the corresponding sub-image is selected.
This procedure keeps on average approximately $99\,\%$ of the total simulated energy.
In a second step (going from the second to the third row in the figure), each energy deposition is crystal-wise divided by the sum of the energy falling into the selected part of the image, and a power-scaling of $E^{p}$ is applied to the normalized crystal energies to reduce the sparsity of the images. 
As in ref.~\cite{Baldi:2020hjm}, choosing $p=0.3$ leads to a notable improvement in our network performance, while other values in the range $\left[0.1,1\right]$ were tested as well in the hyperparameter optimization.

\section{Super resolution network}
\label{sec:SRnetwork}

\begin{figure}[t]
    \includegraphics[width=\textwidth]{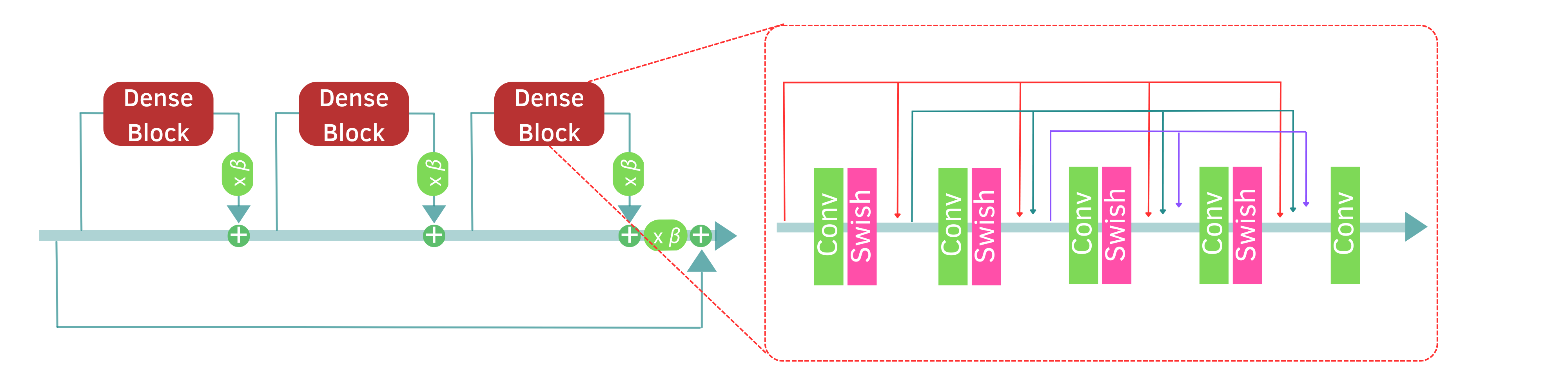}
    \caption{Structure of the RRDB blocks used in this study, consisting of three dense blocks, which each contain five convolutional layers with Swish activation.
    The residual connections are scaled by a free parameter $\beta$.}
    \label{fig:SRnetwork/RRDB}
\end{figure}

A successful application of GANs to the SR task was achieved by the SRGAN~\cite{SISRGAN}. 
It uses a deep convolutional neural network based on residual learning~\cite{DBLP:journals/corr/HeZRS15} as generator and showed the capability of restoring realistic textures with an upsampling factor of four from downsampled LR images with the help of a new perceptual loss term~\cite{Johnson:2016}.
Our network architecture builds upon the architecture of the ESRGAN~\cite{DBLP:journals/corr/abs-1809-00219}.
The ESRGAN is an enhanced version of the SRGAN, which uses a relativistic loss in the discriminator, a more effective perceptual loss and a deeper generator network constructed with residual-in-residual dense blocks (RRDBs) as its fundamental component.
The RRDBs, shown in Fig.~\ref{fig:SRnetwork/RRDB}, consist of three dense blocks~\cite{DBLP:journals/corr/HuangLW16a} connected by residual connections.  
Additionally, a residual connection is used to link the input of the RRDB to its output.
The dense blocks comprise five convolutional layers, where each layer incorporates the outputs of all preceding layers within the block as its inputs.

\begin{figure}
    \includegraphics[width=\textwidth]{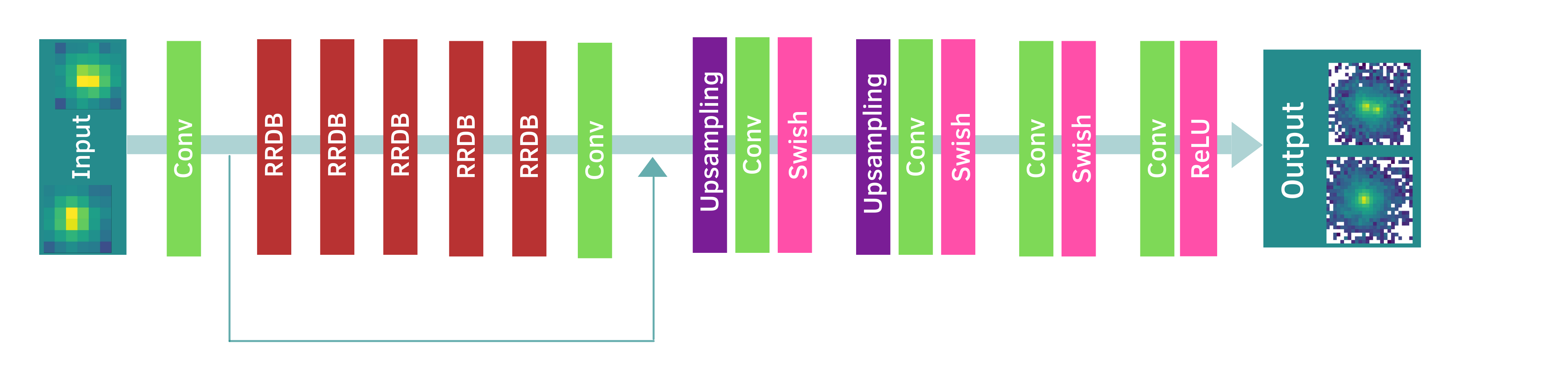}
    \caption{Schematic representation of the generator architecture. 
    The low-resolution input images are fed into a convolutional layer. 
    The extracted features are passed into a block of five RRDBs followed by a convolutional layer. 
    A residual connection adds the output of the first convolutional layer.
    The upsampling takes place in the two upsampling layers, each of which doubles the number of pixels along the $x$- and $y$-axis of the images, which is followed by two additional convolutional layers.}
    \label{fig:SRnetwork/generator}
\end{figure}

The architecture of our generator network is illustrated in Fig.~\ref{fig:SRnetwork/generator}.
The LR input images are first processed by a convolutional layer, after which they are passed through five RRDBs and another convolutional layer to extract high-level features.
The output of this layer is then combined with the output of the first layer via a skip connection~\cite{10.1007/978-3-319-46493-0_38}.
In contrast to the original design, we use Swish~\cite{DBLP:journals/corr/abs-1710-05941} instead of Leaky ReLU as activation functions inside the RRDBs, as this improved the training stability.
The upsampling of the LR images is done with two upsampling blocks, each containing an upsampling layer that doubles the number of pixels along the $x$- and $y$-axes using nearest-neighbor interpolation, followed by a convolutional layer with Swish activation. 
As in the original ESRGAN architecture, two additional convolutional layers are employed after the upsampling blocks, the first is activated using Swish and the latter using ReLU, which avoids the generation of negative energies.

Each convolutional layer in the generator consists of 32 filters with $3\times3$ kernels.
The striding is set to one and zero-padding is used to preserve the resolution of the images when applying convolutions.   
In total, the generator network has around 2.1 million trainable parameters. 

We train the generator to perform realistic upsampling using the Wasser\-stein-GAN (WGAN) approach~\cite{Arjovsky2017WassersteinG}, 
which aims to minimize the \mbox{Wasserstein-1} distance between the probability distributions $\mathcal{P}$ of the real HR images and the generated SR images. 
We can write the Wasserstein distance between these distributions as
\begin{equation}
    W(\mathcal{P}_\text{HR}, \mathcal{P}_\text{SR}) = \sup_{\lvert\lvert f \rvert\rvert_L \leq 1} \left( \mathbb{E}_{x\,\in\, \mathcal{P}_\text{HR}} \left[ f(x) \right] - \mathbb{E}_{\tilde{x}\,\in\, \mathcal{P}_\text{SR}} \left[ f(\tilde{x}) \right] \right)\,, 
    \label{eqn:Wasserstein_distance}
\end{equation}
with $\lvert\lvert f \rvert\rvert_L \leq 1$ denoting the set of Lipschitz continuous functions applied to our calorimeter images and $\mathbb{E}$ denoting the expectation value.
The function $f$ that maximizes the expression in Eq.~\eqref{eqn:Wasserstein_distance} is approximated by training the critic network while at the same time forcing it to fulfill the Lipschitz condition.
Several techniques exist to constrain the critic to be Lipschitz continuous, and we use the gradient penalty (GP) proposed in Ref.~\cite{Gulrajani:2017}. 
The GP introduces an additional term in the critic loss that penalizes the network to obtain gradient norms, with respect to its inputs, that deviate from one.
In this setup, the loss function for a critic network $C$ can be written as
\begin{equation}
    \mathcal{L}_\text{C} = \mathbb{E}_{\tilde{x}\,\in\, \mathcal{P}_\text{SR}} \left[ C(\tilde{x}) \right] - \mathbb{E}_{x\,\in\, \mathcal{P}_\text{HR}} \left[ C(x) \right] + \lambda_\text{GP} \mathbb{E}_{\hat{x}\,\in\, \mathcal{P}_{\hat{x}}} \left[ \, \left( \lvert\lvert \nabla_{\hat{x}} C(\hat{x}) \rvert \rvert_2 - 1  \right)^2 \, \right].
    \label{eqn:critic_loss}
\end{equation}
The last term describes the gradient penalty with strength parameter $\lambda_\text{GP}$ and is calculated along straight lines $\hat{x}$ that are randomly sampled between given pairs of HR images $x$ and SR images $\tilde{x}$ as $\hat{x} = x + \alpha (\tilde{x}-x)$, where $\alpha$ is randomly sampled from a uniform distribution between 0~and~1.

\begin{figure}[t]
    \includegraphics[width=\textwidth]{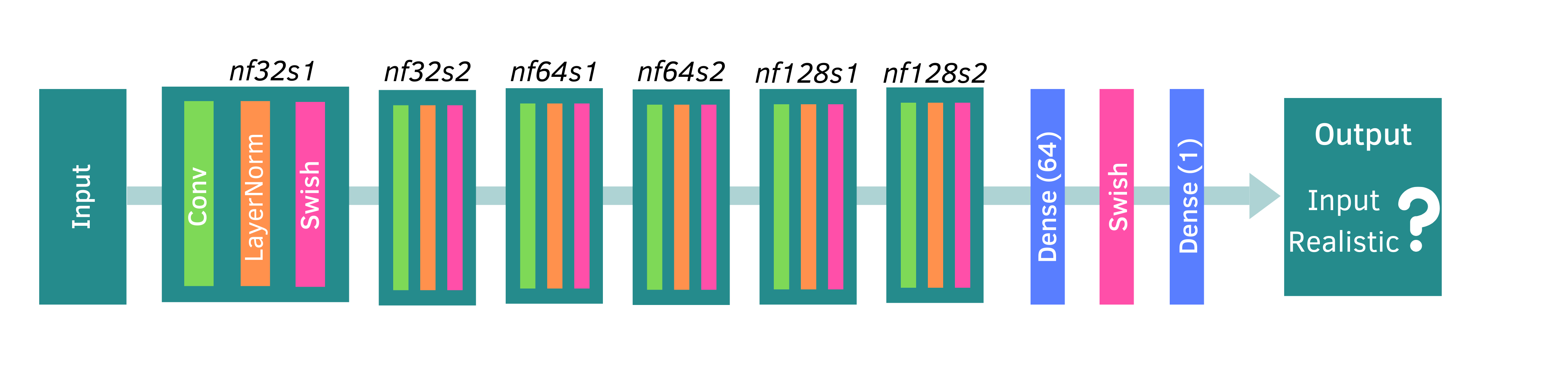}
    \caption{ 
    Illustration of the critic architecture, consisting of six convolutional layers, each followed by Layer Normalization and Swish activation function, and two dense layers. 
    The number of filters \textit{nf} and the striding parameters \textit{s} of the convolutional layers are given, as well as the number of nodes of the dense layers.
    }
    \label{fig:SRnetwork/discriminator}
\end{figure}

The structure of our critic network is shown in Fig.~\ref{fig:SRnetwork/discriminator} and is similar to the discriminators used in the original SRGAN and ESRGAN.
The network receives either HR or SR images as input and outputs a single number discriminating between these image classes.
It consists of six convolutional layers and two dense layers. 
The convolutional layers are placed in an alternating structure with strides of $s=1$ and $s=2$.
Each layer with stride convolutions ($s=2$) halves the dimension in the $x$- and $y$-direction of its input. 
The number of filters is doubled in the third and fourth convolutional layer (64~filters) and again doubled in the fifth and sixth layer (128~filters).
All convolutional layers use $3\times3$ kernels and zero-padding. 
After each convolutional layer, we use Layer Normalization~\cite{Ba2016LayerN}, as recommended in Ref.~\cite{Gulrajani:2017}, instead of the originally proposed Batch Normalization~\cite{ioffe2015batch}, and we use the Swish activation function.
The output of the last convolutional layer is flattened and passed to a dense layer with 64 nodes and Swish activation function, followed by the last layer with a single node.  

In addition to the adversarial loss, which uses the critic's output to improve the generated images, we use the concept of perceptual loss~\cite{Johnson:2016} to train the generator.
In contrast to a crystal-wise comparison of energy depositions between a SR calorimeter image and the reference HR image, the feature representations extracted from a hidden layer of a pre-trained convolutional neural network (CNN) are compared between image pairs. 
The ESRGAN uses the VGG19 network~\cite{Simonyan2014VeryDC} trained on the ImageNet~\cite{Russakovsky2014ImageNetLS} dataset and calculates the Euclidean distance between the features extracted from the last convolutional layer. 
Since our calorimeter images strongly differ from the ImageNet examples, 
we use a CNN trained to separate single-photon from neutral-pion-decay calorimeter images for the perceptual loss.
This network is discussed in more detail in Sec.~\ref{sec:results_analysis}.
Similar to the ESRGAN, we use the features extracted from the last (third) convolutional layer, corresponding to a high-level representation of the input images. 
The generator is hence trained to retain features of the images that are important for the classification as photon or pion.
The full generator loss is the sum of the adversarial loss and the perceptual loss, weighted by the parameters $\lambda_\text{adv.}$ and $\lambda_\text{per.}$,
\begin{equation}
    \mathcal{L}_G = \lambda_\text{adv.} \left(\mathbb{E}_{\tilde{x}\,\in\, \mathcal{P}_\text{SR}} \left[ C(\tilde{x}) \right]\right) + \lambda_\text{per.}\Biggl( \sum_{(x,\,\tilde{x})} (\Phi(x) - \Phi(\tilde{x}))^2 \Biggr),
    \label{eqn:generator_loss}
\end{equation}
where $\Phi$ denotes the feature representations of SR images~$\tilde{x}$ and HR images~$x$.

\section{Network training}
\label{sec:training_results}

The super-resolving GANs are trained using \mbox{100,000} photon and \mbox{100,000} neutral pion images.
We adapt several recommendations from Ref.~\cite{Gulrajani:2017} for the training of the WGAN:
we use the Adam optimizer~\cite{kingma2017adam} with learning rate $10^{-4}$ and decay parameters $\beta_1=0$ and $\beta_2=0.9$ and train the critic for five mini-batches before training the generator for one mini-batch. 
We use a batch-size of 32.
In the $20\,\text{GeV}$ setup, the perceptual loss is scaled by $\lambda_\text{per.}=3\cdot10^{-2}$, while $\lambda_\text{per.}=3\cdot10^{-1}$ is used for the $50\,\text{GeV}$ network.
The adversarial term of the generator loss is scaled by $\lambda_\text{adv.}=10^{-5}$.
The critic networks are trained with a gradient-penalty strength of $\lambda_{\text{GP}}=1$.
The networks are implemented using TensorFlow \texttt{2.10.0}~\cite{tensorflow_developers_2022_7604243} and trained for approximately 10 days on an NVIDIA A40 GPU.

The hyperparameters are optimized in a grid-search as follows:
in a first step, the capacities of the networks are varied, in particular the number of RRDBs in the generator.
At the same time, different values for the scaling parameters of the generator and critic loss terms, $\lambda_\text{adv.}$ and $\lambda_{\text{GP}}$ are studied.
These parameters are fixed to the above mentioned values taking in particular the training stability and convergence together with the visual quality of the SR images into account. 
In order to decrease the complexity of the hyperparameter optimization, the perceptual loss is not included in these first studies, i.e. $\lambda_\text{per.}=0$ is used. 
The performance depends only marginally on the generator capacity in the tested range of 1--10 RRDBs, hence  an intermediate value of 5 is chosen.
The smaller dimension of our HR and SR images requires a reduction of the number of convolutional layers in the critic compared to the architecture used in the original ESRGAN from eight to six, since the layers with strided convolutions ($s=2$) each halve the number of pixels along both image axes. 
In addition, the number of nodes in the first dense layer in the critic is reduced from 1024 to 64, which significantly reduces the training time while no differences in the performance are found.
With this setup, the GAN trainings run stably for both particle energies and produce realistic SR images where no obvious artefacts are observed.

\begin{figure}[h!t]
    \centering
    \subfloat{\includegraphics[width=0.42\textwidth]{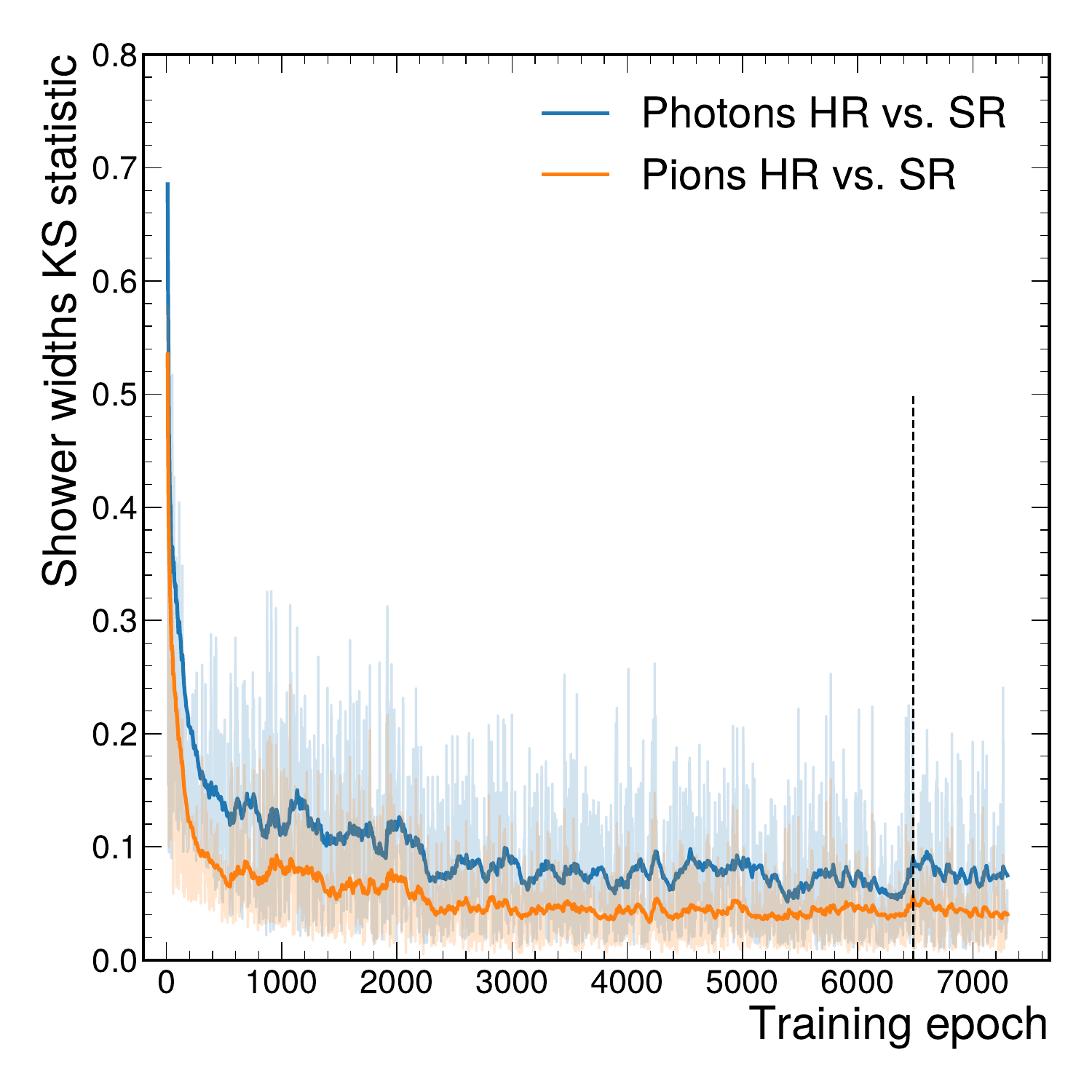}}
    \subfloat{\includegraphics[width=0.42\textwidth]{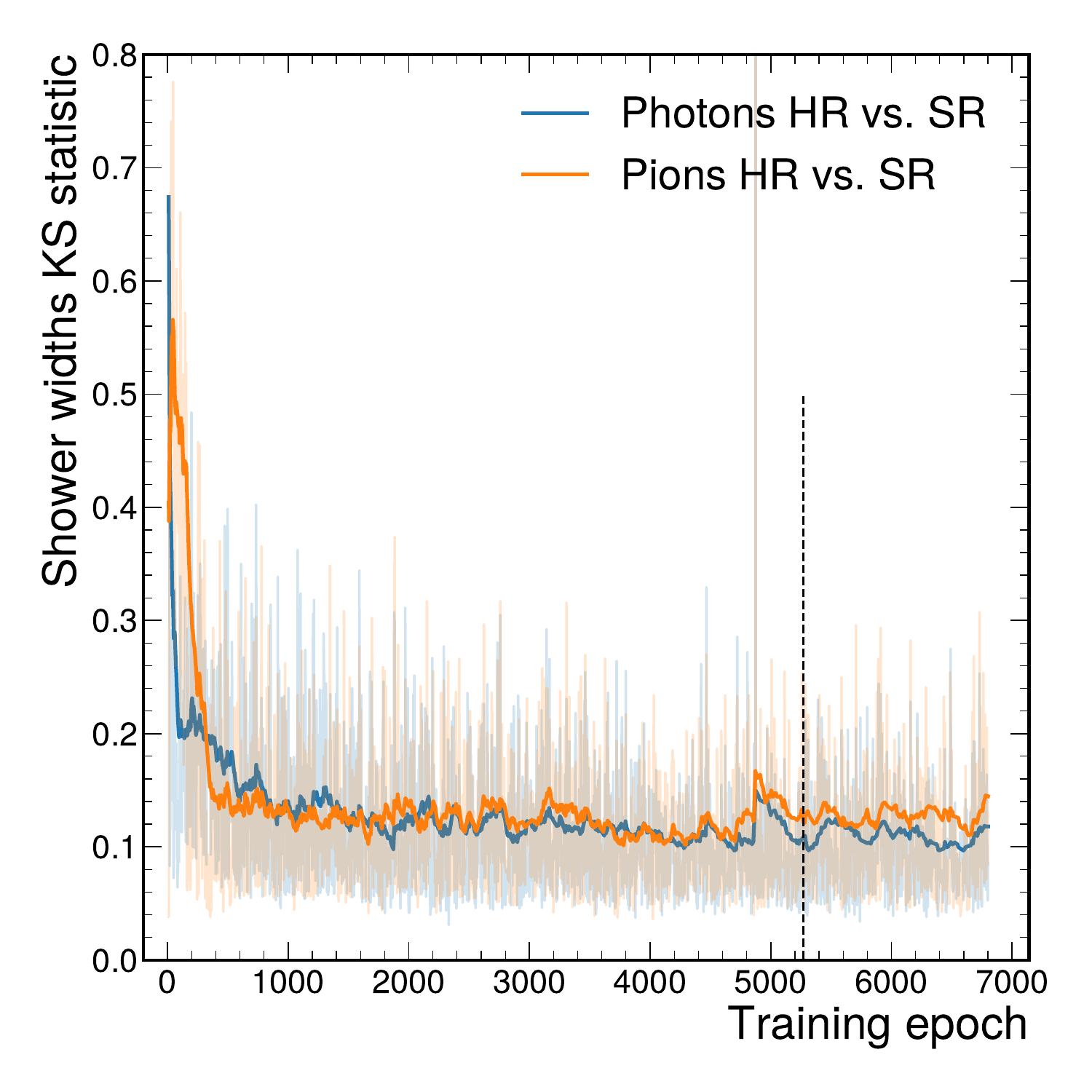}}
    \caption{
    Kolmogorov-Smirnov statistic calculated between the HR and SR shower width distributions as a function of the training epoch, separately for photons and pions.
    The $20\,\GeV$ setup is shown on the left, the $50\,\GeV$ setup is shown on the right.
    The stronger lines are obtained by smoothing the original values shown with the lighter colors.
    We select the epoch where the mean of the photon and pion statistics (without smoothing) is at its minimum, indicated by the black dashed line.
    }
    \label{fig:results/shower_width_history}
\end{figure}

In a second step, the perceptual loss is included in the training with the particular goal to penalize the generator for confusing the two particle types. 
To evaluate and optimize its impact, we monitor the capability of the CNNs pre-trained on the HR images to distinguish between the SR photon and pion examples and analyze the impact on shapes of the electromagnetic shower and the differences between photons and pions.
We determine the distribution of the shower width in the SR images and compare it to the distribution obtained from the HR images. 
In LHC experiments, similar variables describing the shower shape are used to discriminate between photons and other signatures from hadronic activity~\cite{ATLAS_EGamma,CMS_EGamma}.
We define the width of a shower image with crystal indices $i$ as
\begin{equation}
    W = \frac{\sum_{i} \Delta R_i E_i}{\sum_{i} E_i},
\label{eq:width}
\end{equation}
where $E_i$ denotes the energy measured in a crystal and $\Delta R_i$ is its angular distance to the barycenter of the shower in units of rad.
We obtain the distributions separately for photons and pions and monitor the Kolmogorov–Smirnov (KS) statistic between each HR and SR width distribution during the training. 
The values obtained for the KS statistics are shown in Fig.~\ref{fig:results/shower_width_history}.
The epoch with the lowest mean of the KS statistic for pions and photons is finally selected.
Since the perceptual loss uses individual CNNs in the $20\,\text{GeV}$ and $50\,\text{GeV}$ setups, different values of the corresponding relative weight ($\lambda_\text{per.}$) are found to yield the best performance.
We observe that including this additional loss term with the optimized weight improves the pion rejection\footnote{The rejection is defined as the inverse of the efficiency, i.e. $1 /\!\left(\text{false positive rate}\right)$.} obtained from the pre-trained CNNs applied to the SR images compared to trainings without perceptual loss by up to a factor of five, depending on the photon identification efficiency.

\begin{figure}[h!]
    \centering
    \subfloat{\includegraphics[width=0.5\textwidth]{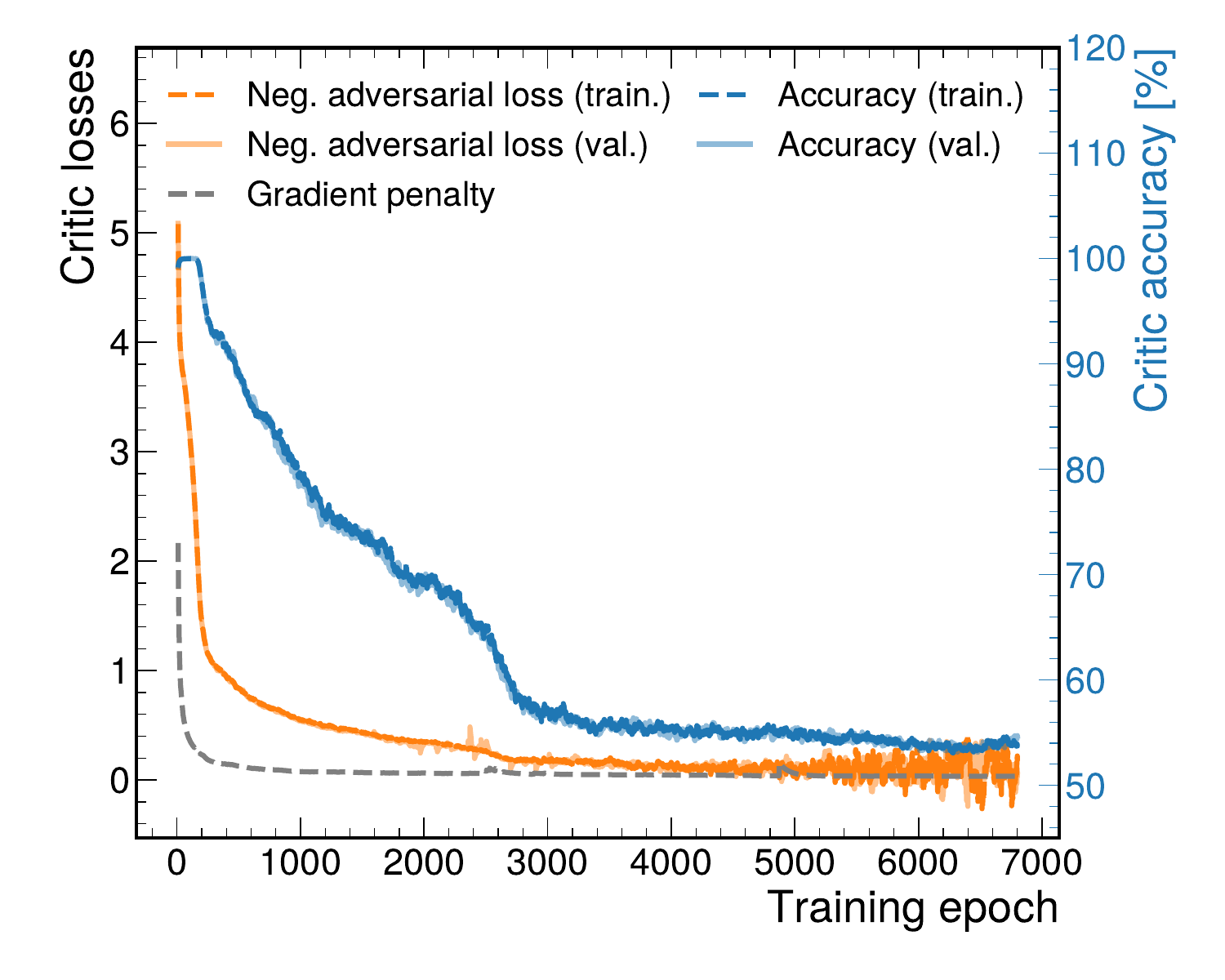}}
    \subfloat{\includegraphics[width=0.5\textwidth]{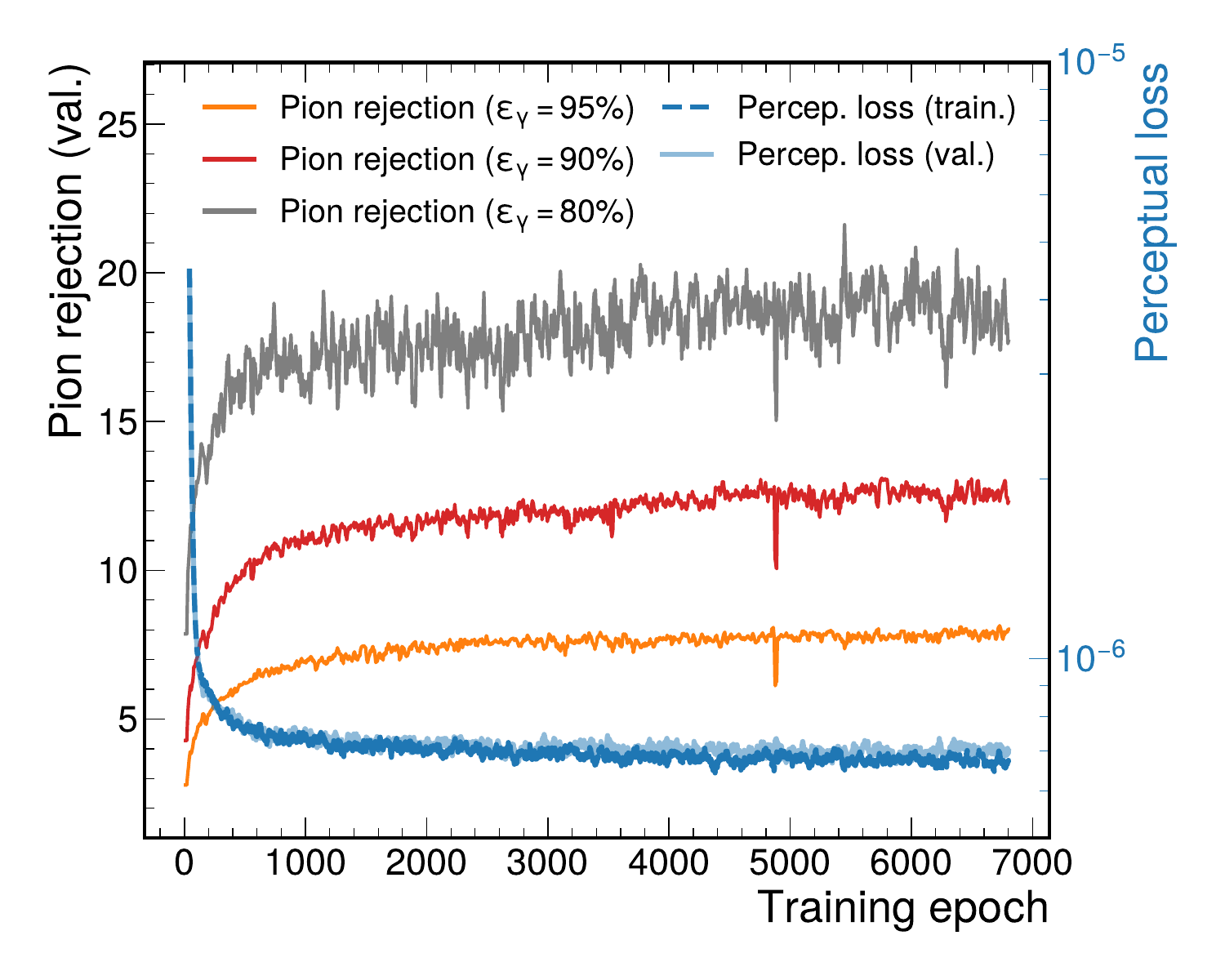}}
    \caption{Different parts of the loss functions and metrics during the training of the $50\,\text{GeV}$ network, where ``train.'' (``val.'') refers to loss/metrics evaluated on the training (validation) sample.
    Left: losses of the critic network and its accuracy in discriminating between HR and SR images.
    Right: perceptual loss used for the generator training and the pion rejection at several fixed photon efficiencies obtained with the pre-trained CNN.
    }
    \label{fig:results/GAN_history}
\end{figure}

In Fig.~\ref{fig:results/GAN_history}, the evolution of the different parts of the loss functions during training as well as several metrics are shown for the example of the $50\,\text{GeV}$ network.
At the start of the training, the critic network is able to discriminate between the original HR and the generated SR images with an accuracy of $100\,\%$.
It can be seen that during the training, the critic accuracy approaches a value slightly above $50\,\%$, while the critic loss---which approximates the Wasserstein distance---tends towards zero.
In addition, the evolution of pion rejections for fixed values of the photon efficiency is shown, which is evaluated on SR images with the CNN that was pre-trained on HR images.
The pion rejections increase as the perceptual loss decreases.

\begin{figure}
    \centering
    \includegraphics[width=\textwidth]{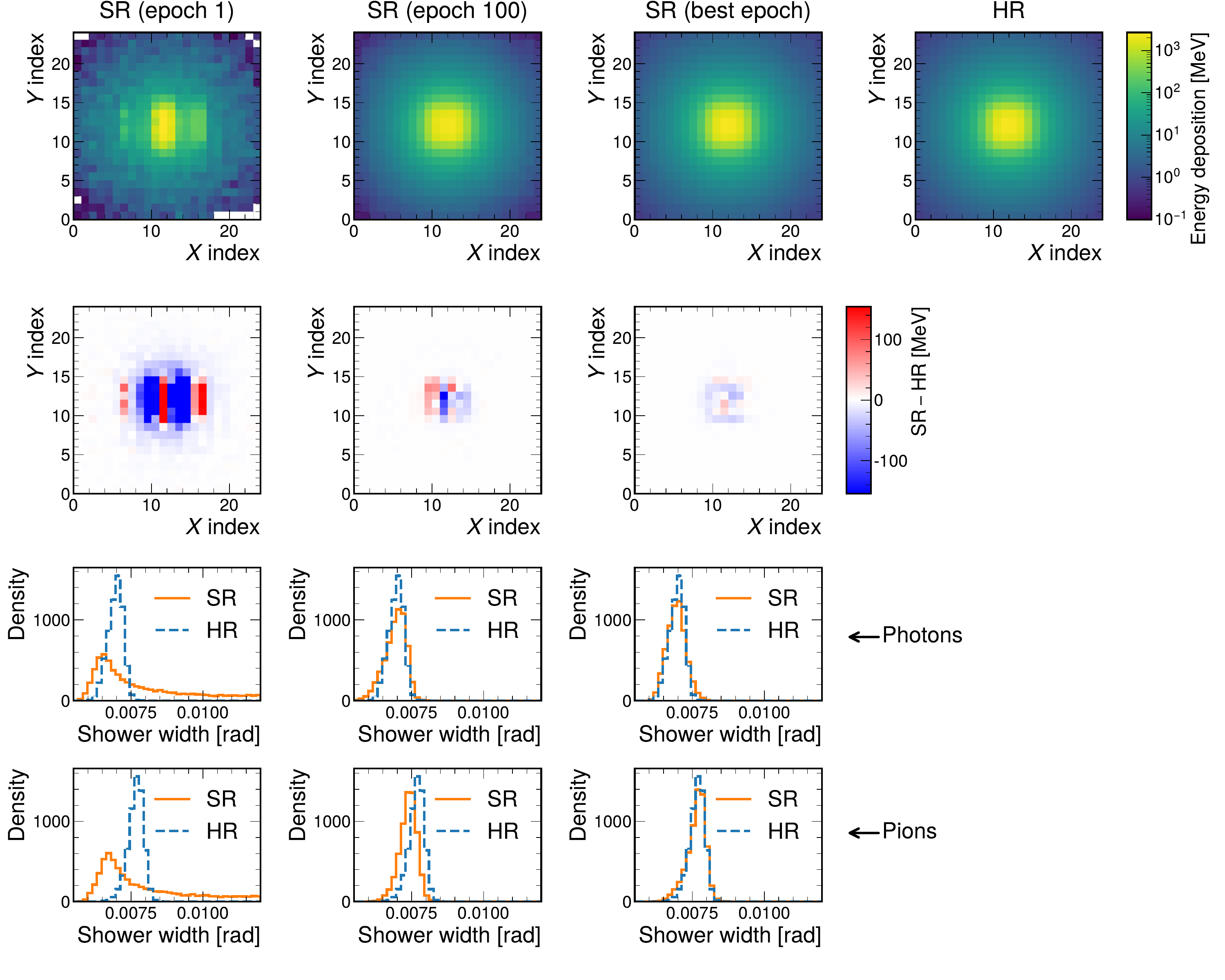}
    \caption{
    Evolution of the image quality during the training of the $50\,\GeV$ network.
    The top row shows the average across all photon and pion images in the validation sample.
    From left to right, the average SR image after one epoch, after 100 epochs, at the selected best epoch, and the simulated HR average are displayed.
    The second row from the top presents the difference between these SR images and the HR image.
    In the third and bottom row, the corresponding SR shower widths are compared to the HR shower widths for photons and pions, respectively.}
    \label{fig:results/training_progress}
\end{figure}

The training progress is also visualized in Fig.~\ref{fig:results/training_progress}.
In the initial stages of the training, distinct artefacts are evident in the SR images.
By averaging over all images, biases in the spatial distribution of the predicted energy depositions become visible, which largely disappear after around 100 training epochs.
Similarly, the network learns to generate photons and pions with shower widths almost matching the HR distributions within these initial 100 epochs.
However, we still observe improvements in the generated widths and in other metrics like the critic accuracy or pion rejections up to around \mbox{5,000} training epochs.

\FloatBarrier

\section{Results}
\label{sec:results_analysis}

\begin{figure}[p]
    \centering
    \subfloat{\includegraphics[width=0.9\textwidth]{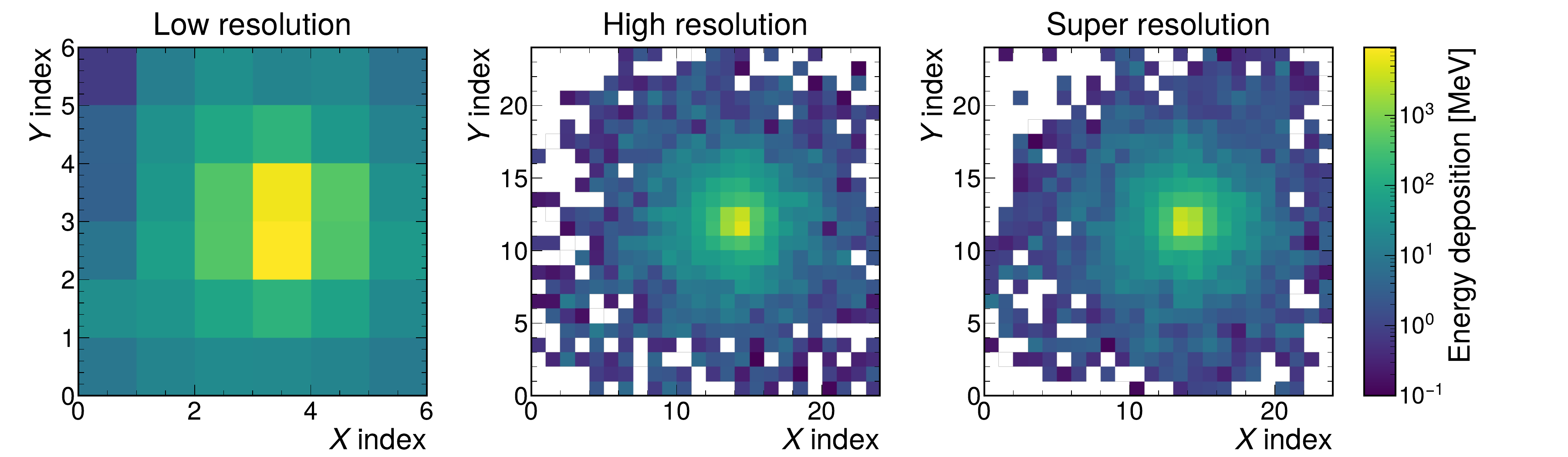}}\\
    \subfloat{\includegraphics[width=0.9\textwidth]{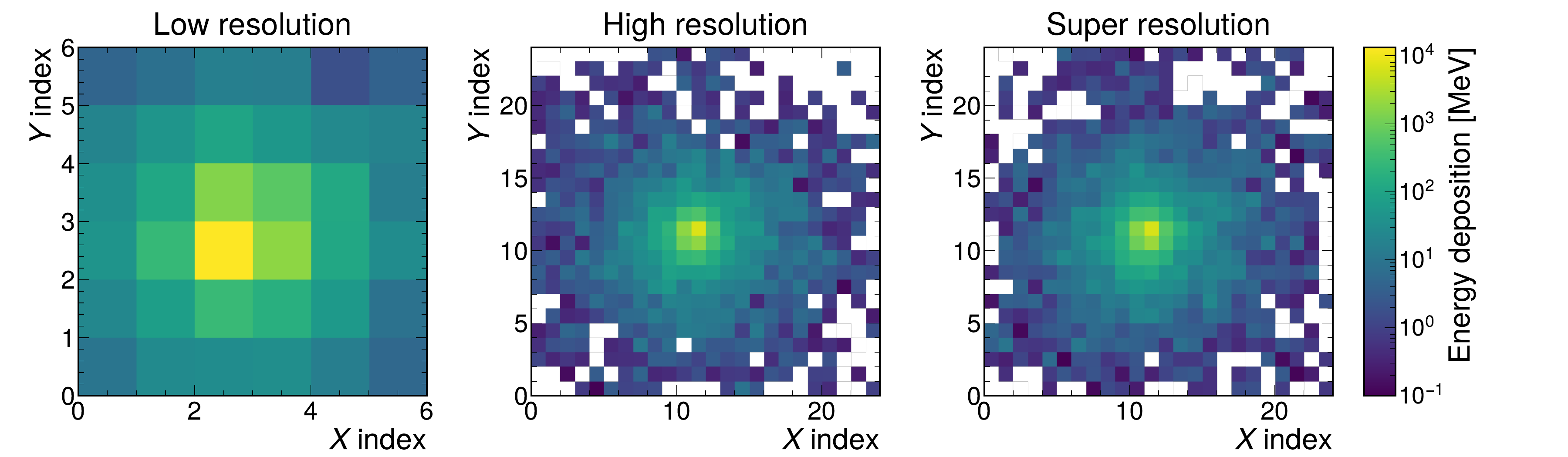}}\\
    \subfloat{\includegraphics[width=0.9\textwidth]{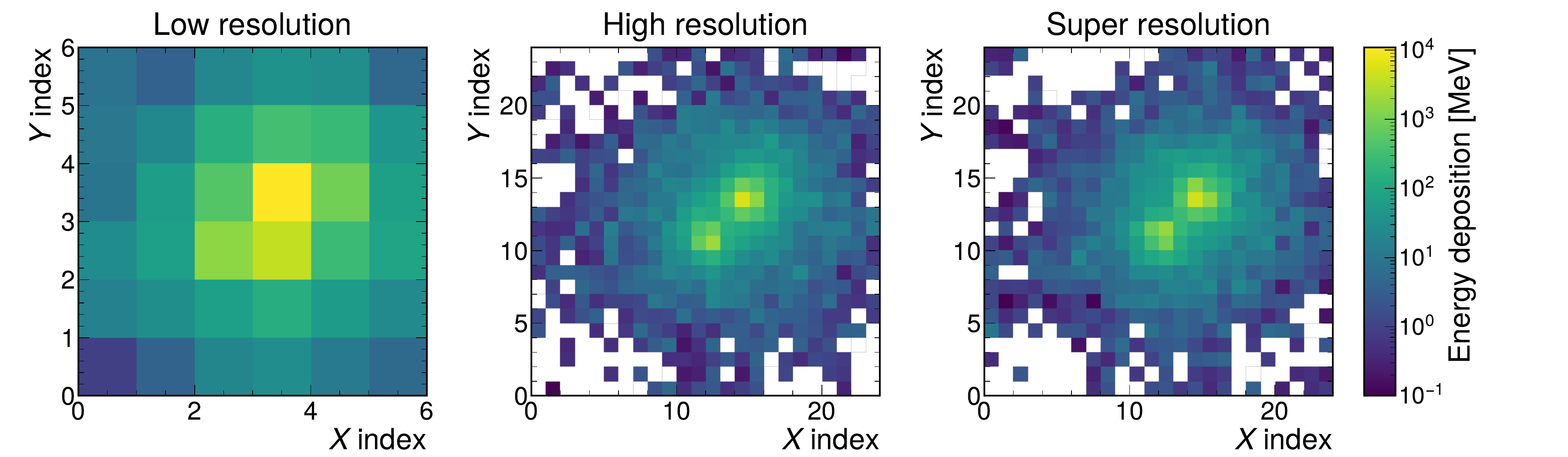}}\\
    \subfloat{\includegraphics[width=0.9\textwidth]{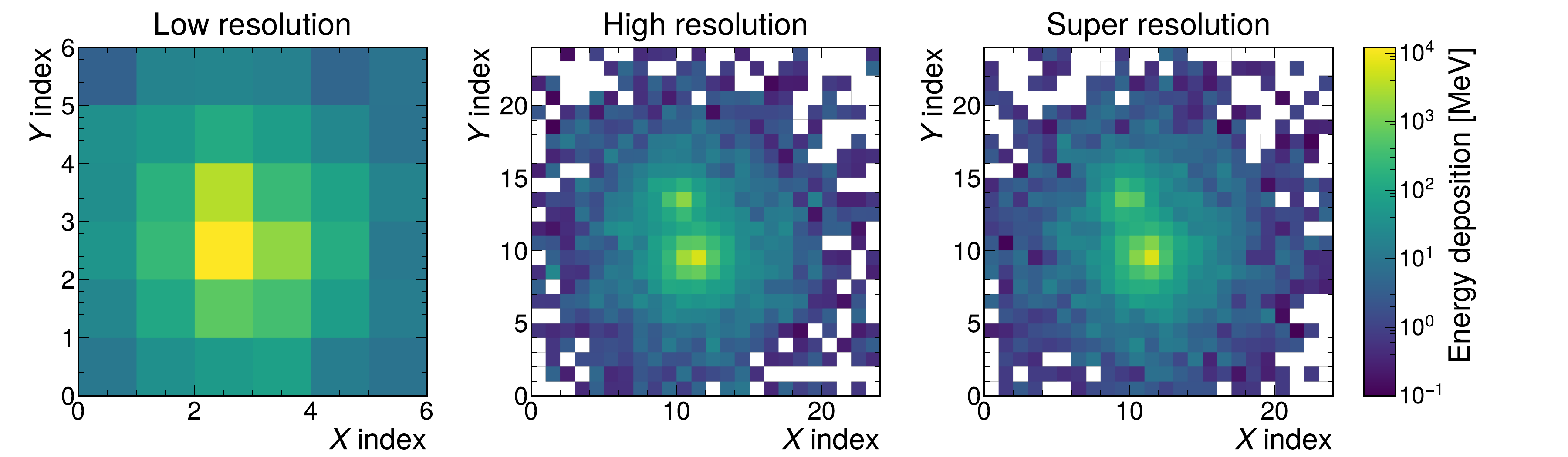}}
    \caption{Example SR images (right column) with their corresponding LR (left) and HR (middle) versions for the $20\,\GeV$ network. 
      The first two rows show photon examples, the bottom two rows show pion examples.
    }
    \label{fig:results/example_images_20GeV}
\end{figure}

\begin{figure}[p]
    \centering
    \subfloat{\includegraphics[width=0.9\textwidth]{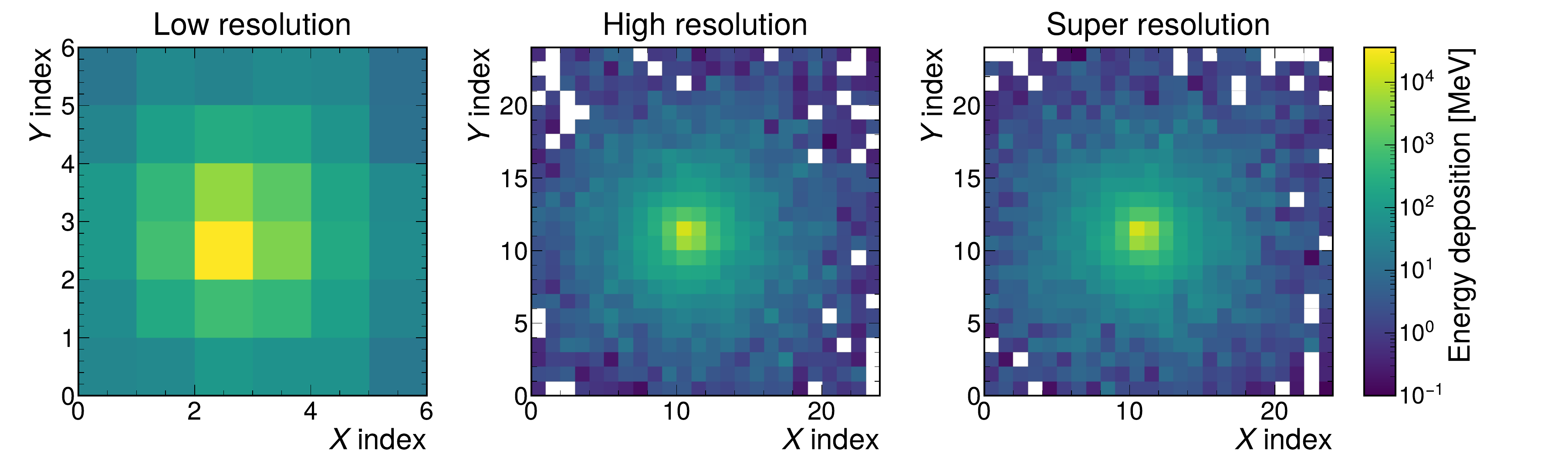}}\\
    \subfloat{\includegraphics[width=0.9\textwidth]{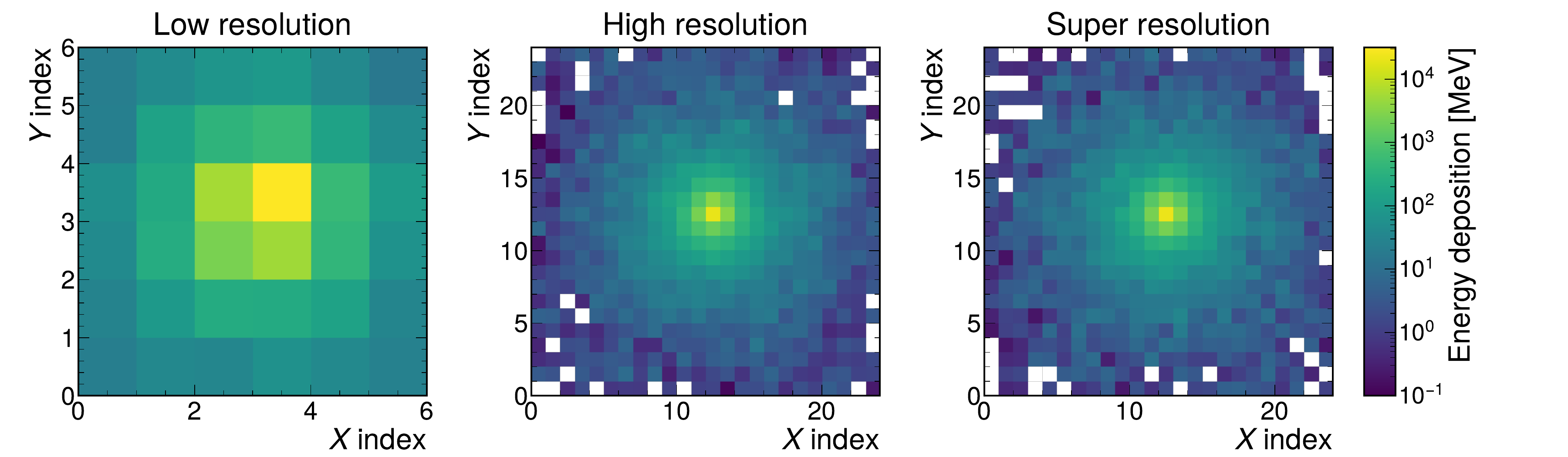}}\\
    \subfloat{\includegraphics[width=0.9\textwidth]{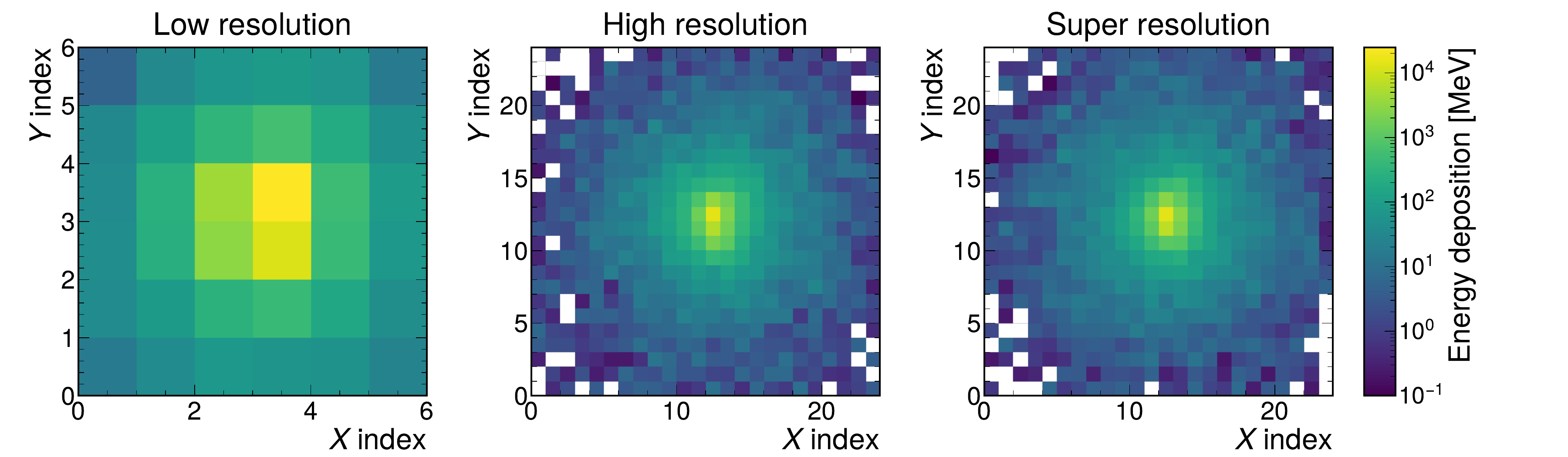}}\\
    \subfloat{\includegraphics[width=0.9\textwidth]{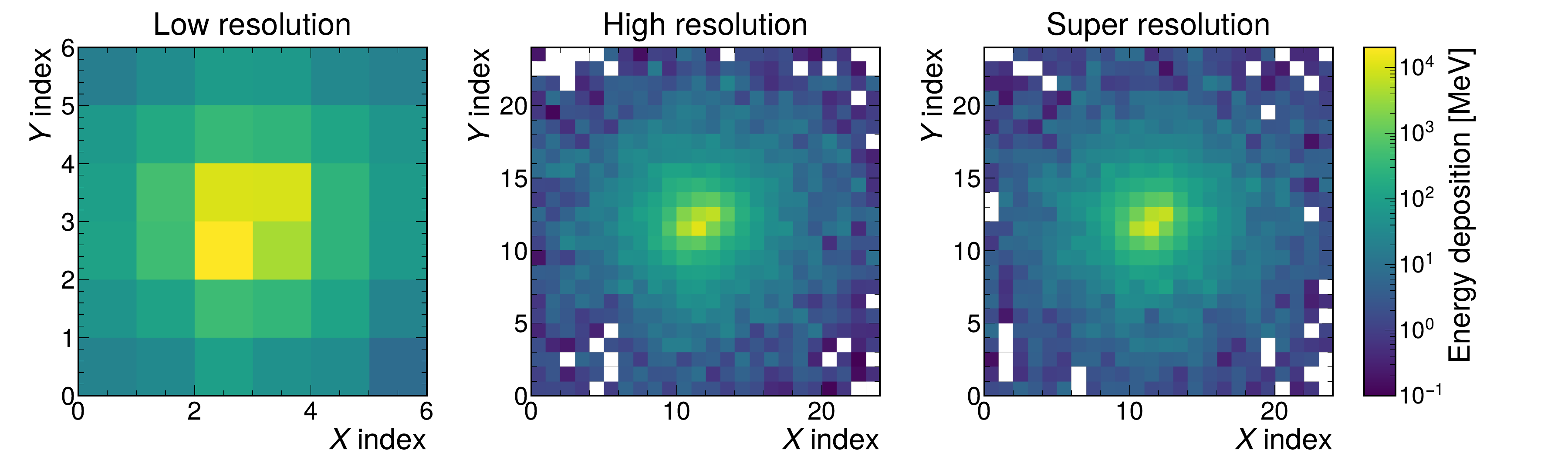}}
    \caption{Example SR images (right column) with their corresponding LR (left) and HR (middle) versions for the $50\,\GeV$ network. 
      The first two rows show photon examples, the bottom two rows show pion examples.
    }
    \label{fig:results/example_images_50GeV}
\end{figure}

After training the SR networks, we study the properties of the upsampled images and discuss possible use cases at hadron-collider experiments.
Example predictions of the generator network are shown in Fig.~\ref{fig:results/example_images_20GeV} for the $20\,\text{GeV}$ network and in Fig.~\ref{fig:results/example_images_50GeV} for the $50\,\text{GeV}$ network, respectively.
For each energy, two randomly picked examples for each particle type are included, comparing the LR image, which was passed to the SR network, to the corresponding HR image and the generated SR version.
In general, we observe that the obtained SR images have a high perceptual similarity with the HR simulation. 

Typically, the main visual properties of the HR images are also found in the generated SR versions.
In particular, we find clear single peaks in the SR photon images and typically two distinct peaks in the pion SR images. 
Furthermore, the position and orientation of these peaks often matches the one of the simulated HR images well, although this information is often difficult to extract from the LR images by eye.

The main difference between the $20\,\text{GeV}$ and $50\,\text{GeV}$ examples is the angle between the photons from the pion decays. 
Comparing the pion examples in Fig.~\ref{fig:results/example_images_20GeV} and Fig.~\ref{fig:results/example_images_50GeV}, the $20\,\text{GeV}$ pions appear as a single merged shower in the LR calorimeter, while they are well resolved as two photons in the HR calorimeter. 
However, asymmetries in the LR calorimeter pion images allow the SR network to generate separate peaks in SR images that often coincide with the peaks in their HR counterparts.
The decay products of the $50\,\text{GeV}$ pions typically appear as two overlapping showers even in the HR calorimeter.
Also in the case of these merged showers, the SR network often reproduces the main features of the HR images.

\begin{figure}[t]
    \centering
    \subfloat{\includegraphics[width=0.5\textwidth]{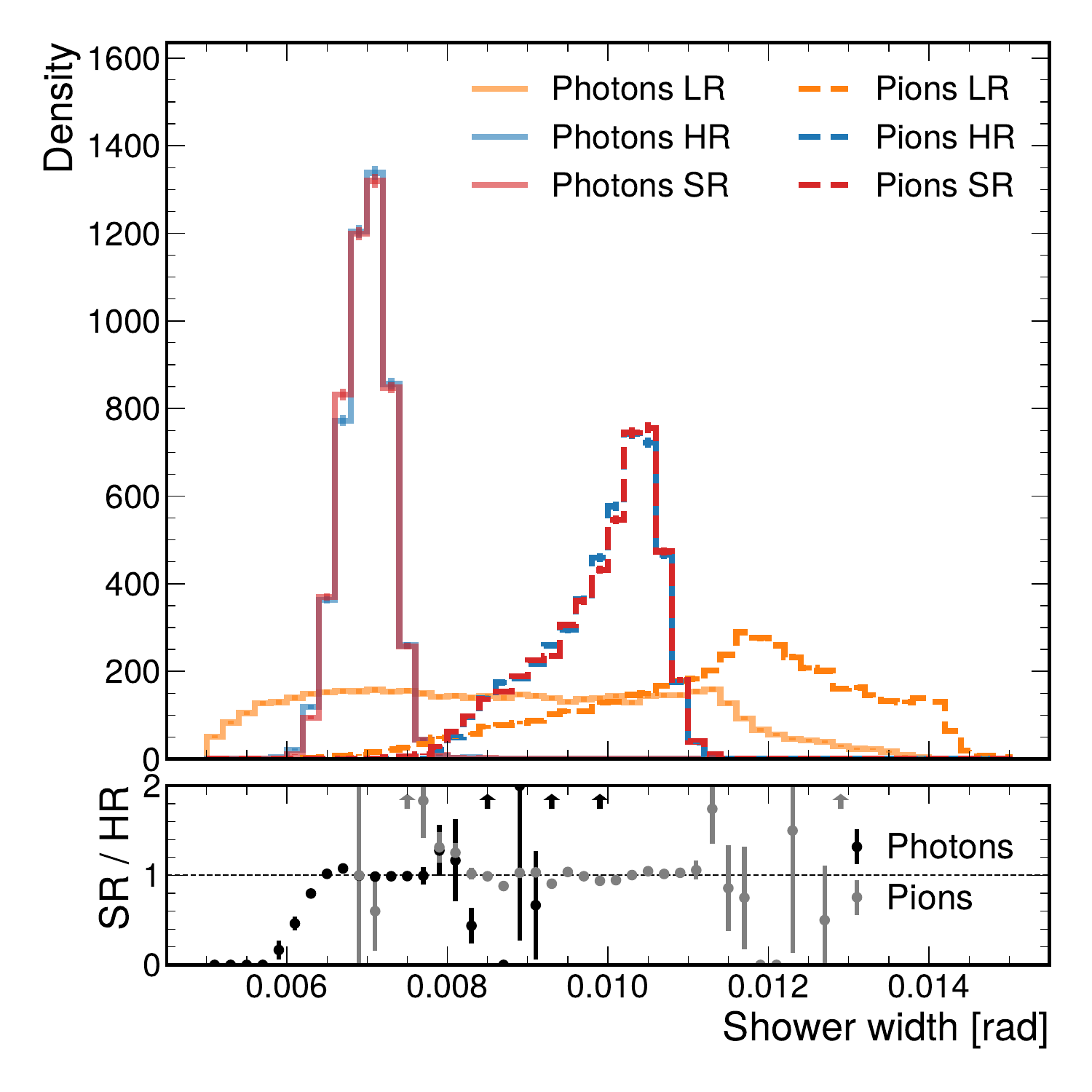}}
    \subfloat{\includegraphics[width=0.5\textwidth]{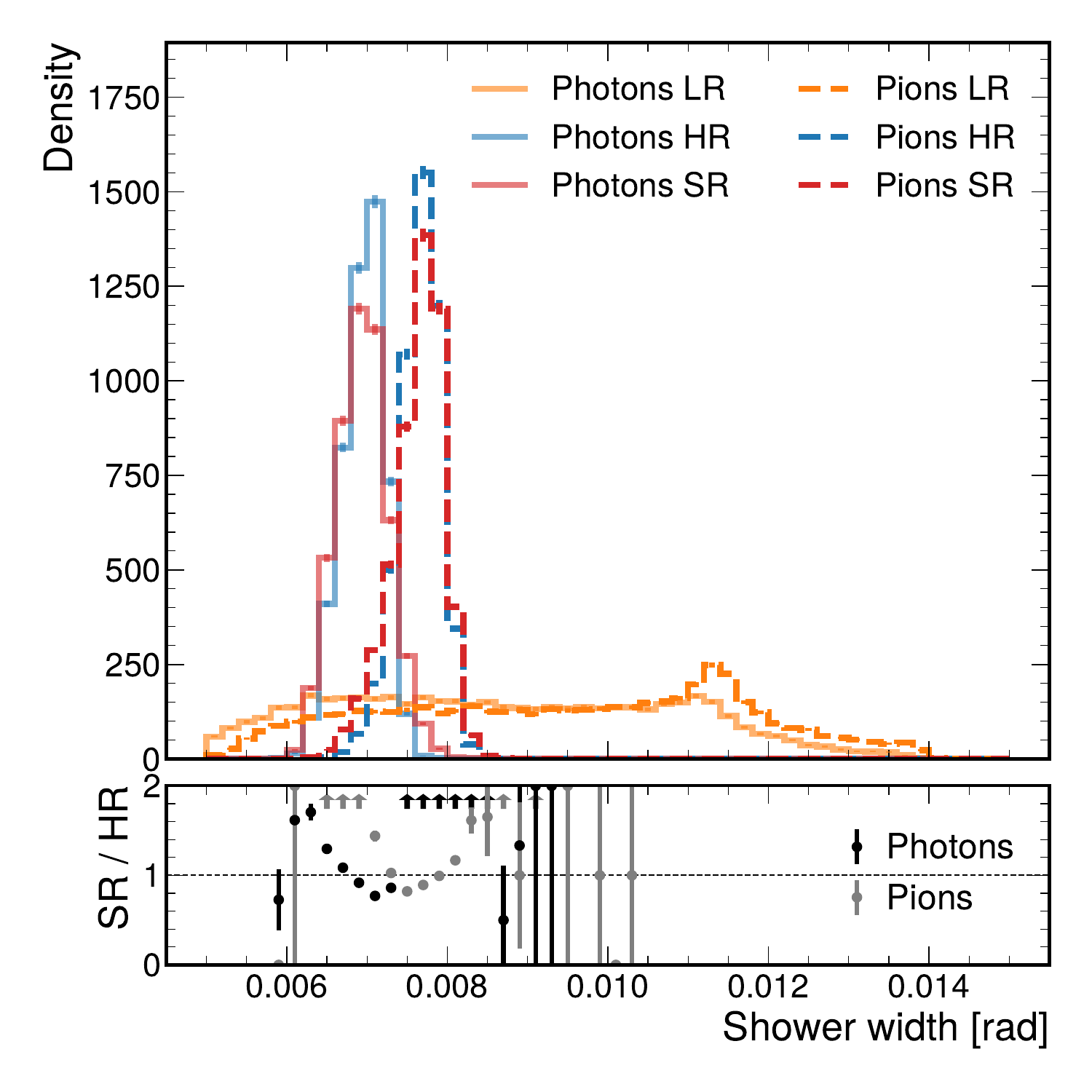}}
    \caption{Normalized distribution of the shower widths for the $20\,\GeV$ particles (left) and for the $50\,\GeV$ particles (right).
      Pion shower widths are shown with the dashed lines, while the solid lines show the photon distributions.
      In addition, the ratio of the SR and the HR distribution is shown.
      Arrows indicate that the value is out of the chosen $y$-axis range of the ratio plot.
      The error bars indicate the statistical uncertainties.
    }
    \label{fig:results/shower_widths}
\end{figure}

As an example of a ``shower-shape variable'', which are often used as features in photon identification algorithms at LHC experiments, we show the shower width in Fig.~\ref{fig:results/shower_widths}, as defined in Eq.~\eqref{eq:width}.
For the $20\,\text{GeV}$ particles, the LR calorimeter can resolve significant differences between photon and pion shower widths, however, with a binning as in Fig.~\ref{fig:results/shower_widths}, the fraction of overlapping area between the photon and pion width histograms is around $52\,\%$.
Comparing to the corresponding HR distributions, it is clearly visible that the higher spatial resolution allows for a better measurement of this quantity.
Hence, shower-shape variables have a much better power to discriminate between photons and pions with the HR calorimeter.
The fraction of overlapping area reduces in the HR histograms to approximately $0.53\,\%$.
Although we train our SR networks on mixed datasets containing photon and pion examples, the shower width distribution obtained from the SR photons and pions closely follow the HR distributions.
Here, the overlapping area is around $0.90\,\%$ and thus heavily reduced compared to the LR case. 
At $50\,\text{GeV}$, the LR width distributions for photons and pions become more similar and the overlapping area increases to $85\,\%$. 
Here, the typical distance between the two photons from the pion decays is much smaller than one crystal width. 
Also in the HR calorimeter, the width distributions appear closer together, but this variable still provides a good separation with an overlap of around $19\,\%$.
The SR distributions match the HR widths less precisely than in the $20\,\text{GeV}$ case, because the discrimination of the classes is more difficult.
However, the overlapping area of around $29\,\%$ is still much lower than in the LR case. 
Thus for both energies, the separation between photons and pions that can be achieved by such a shower shape variable is significantly improved by using the SR image.

\begin{figure}[t]
    \centering
    \subfloat{\includegraphics[width=0.45\textwidth]{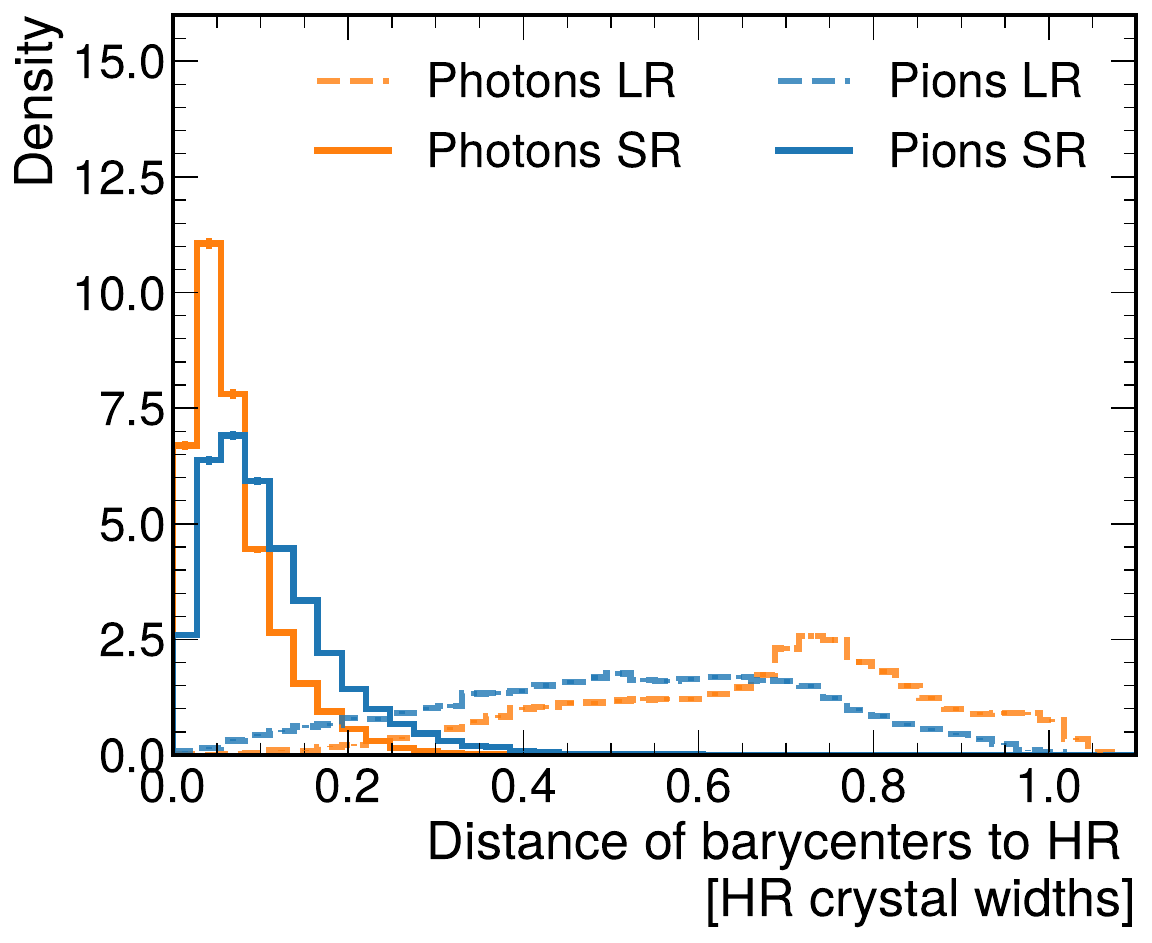}}
    \subfloat{\includegraphics[width=0.45\textwidth]{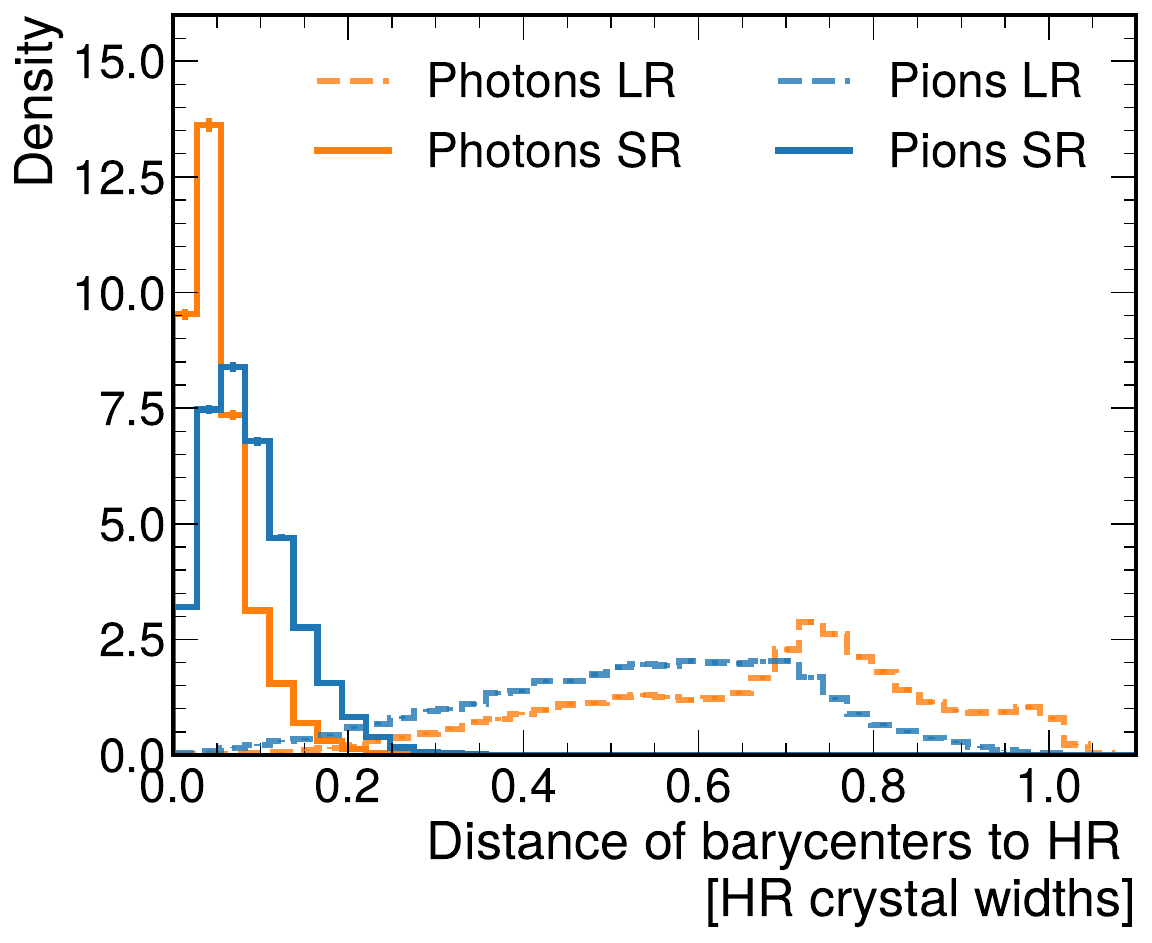}}
    \caption{Normalized distribution of the distance of the barycenter positions of the SR and LR showers from the HR barycenter in units of crystal widths, for the $20\,\GeV$ (left) and $50\,\GeV$ (right) cases.
    }
    \label{fig:results/barycentres}
\end{figure}

In addition to the identification of photon candidates, the measurement of the photon position is a crucial step in the reconstruction chain.
Often, the barycenter position of the cluster of energy depositions is determined and taken as the photon positions' estimate.
The precision in the localization of the barycenter is limited by the granularity of the calorimeter and is important, for example, for the resolution of invariant masses of diphoton decays, such as $H\to\gamma\gamma$. 
To study the effect of the SR technique on the localization of showers, we compare the distances of the barycenter positions of either the SR or LR images and the barycenters of the HR images in Fig.~\ref{fig:results/barycentres}.
We observe that the localization of the photons and pions is significantly improved in SR compared to LR.
From the HR simulation, the generator learns realistic interpolations between the crystals and this leads to an improved determination of the position. 
The actual impact of an improved localization of the photons on the invariant mass resolution of diphoton decays in an experiment depends on further quantities, which we cannot evaluate in our simplified setup, such as the energy resolution of the individual photons and the resolution in the determination of the position of the primary vertex~\cite{ATLAS_HggVertex,CMS_HggVertex}.

Since we observe that differences between the photon and pion images are more prominent in SR than in LR, we study the impact of using SR as a pre-processing step before training classifiers to separate real single photons from fakes induced by neutral-pion decays.
We train CNNs on a dataset of \mbox{100,000} examples, half photons and half pions, which are independent from the samples used for the GAN training.

The CNNs have a comparably simple architecture, beginning with three convolutional layers consisting of 32 filters with a kernel size of $3\times3$.
In these layers, a stride of one and zero-padding are used to conserve the lateral dimensions of the image.
For the HR and SR case, we place a max-pooling layer after each of these layers, which halves the number of pixels in the $x$- and $y$-direction. 
In the LR case, we use only one max-pooling layer after the last convolutional layer and leave out the ones after the first and the second convolutional layer, while the remaining structure is the same as in the HR and SR CNNs.
The output of the last layer is flattened and fed to a dense layer with 10 nodes and ReLU activation, followed by a dense layer with a single node activated by the sigmoid function.
The number of trainable parameters is identical for the CNNs used for the HR or SR images and the LR images.
We train the CNNs using the Adam optimizer with an initial learning rate of $10^{-3}$ and with the binary cross-entropy as loss function.
The trainings are stabilized using L2 regularization with strength of $\mathcal{O}(10^{-4})$, where the exact values are chosen in each training to achieve the best network performance. 
The CNNs trained on the HR images are those that are also used as ``pre-trained CNNs'' for the perceptual loss term in the GAN training.

\begin{figure}
    \centering
    \subfloat{\includegraphics[width=0.5\textwidth]{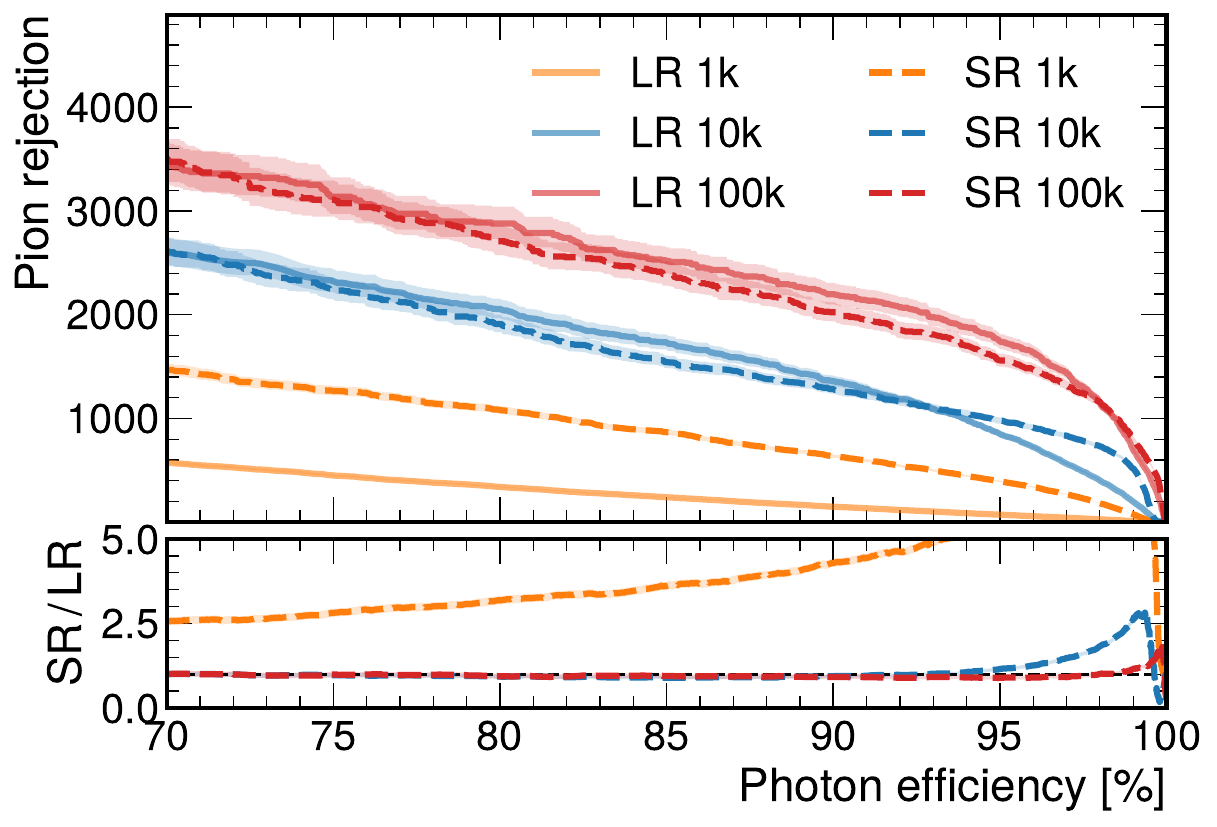}}
    \subfloat{\includegraphics[width=0.5\textwidth]{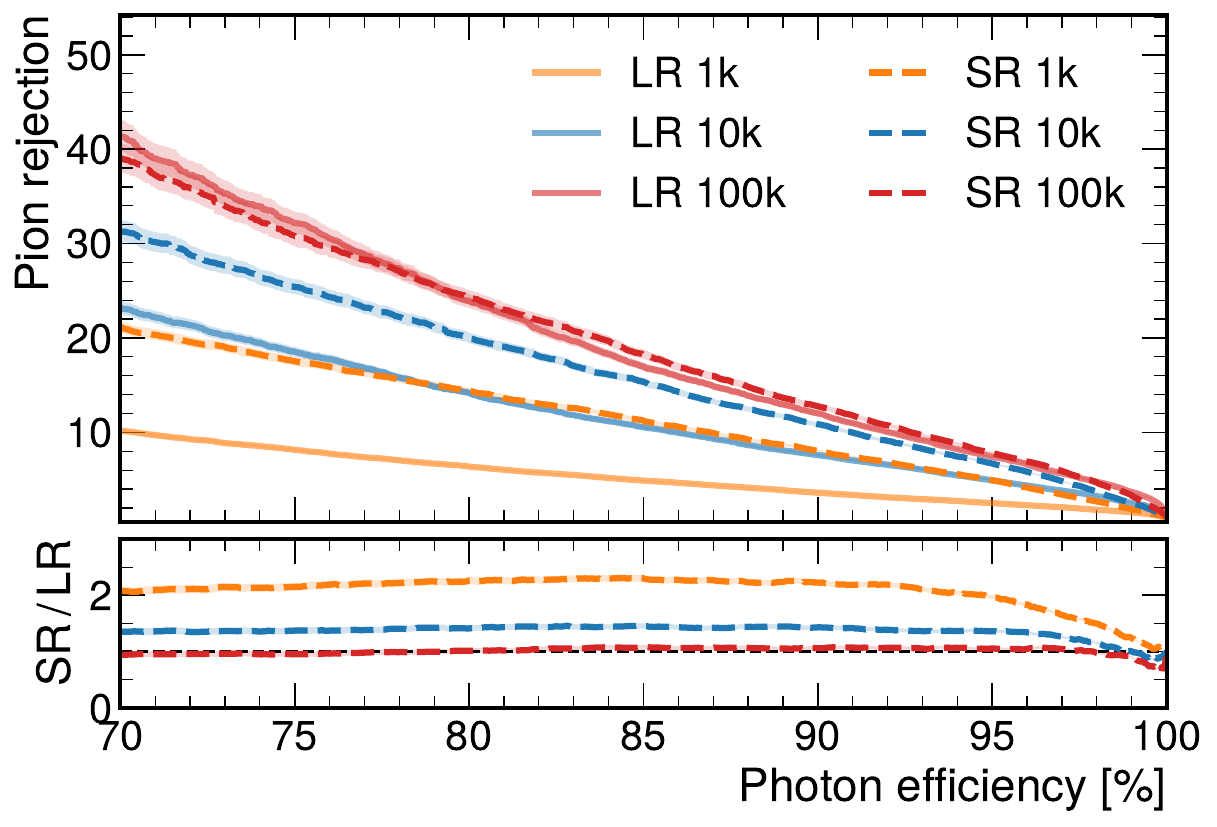}}
    \caption{Classification performance for CNNs trained on either LR or SR images for trainings using different numbers of samples (1k, 10k, 100k).
      The pion rejection is shown as a function of the photon efficiency, for the $20\,\text{GeV}$ (left) and the $50\,\text{GeV}$ simulation (right).
      In addition, the ratio of the SR and the LR pion rejections is shown.
      The error bands represent the statistical uncertainty in the pion rejections.
    }
    \label{fig:results/ROC_curves}
\end{figure}

As expected from the opening angle distributions of the photon from the pion decays (Fig.~\ref{fig:simulation/opening_angles}), large differences are found between the $20\,\text{GeV}$ and the $50\,\text{GeV}$ setups for the separation of photons from pions.
CNNs trained on $20\,\text{GeV}$ images have tiny failure rates in the classification task.
For a given photon efficiency, the pion rejections factors achieved by the $20\,\text{GeV}$ CNNs are two orders of magnitude higher than in the $50\,\text{GeV}$ case.
Comparing the CNNs trained on SR images with the ones trained on LR images, we observe that differences arise depending on the number of samples available for the CNN training.
This is illustrated in Fig.~\ref{fig:results/ROC_curves}, which shows the discrimination achieved by CNNs trained on either the full set of \mbox{100,000} samples or reduced sets of \mbox{10,000} and \mbox{1,000} samples.
The evaluation is done on independent test datasets, which were not used for the GAN or CNN trainings.\footnote{We deploy \mbox{50,000} samples in the $50\,\text{GeV}$ setup, equally photons and pions, but increase the dataset to \mbox{1,000,000} pions and \mbox{100,000} photons in the $20\,\text{GeV}$ setup, because otherwise the statistical uncertainty in the pion rejections is large due to the high rejection values.}
When training the CNN on small datasets, we observe notable improvements when SR is used to enhance the training data.
For both energies, an improvement by a factor of two or more is found in the achieved pion rejections for the case of \mbox{1,000} training samples, over a wide range of photon efficiencies.
In the setup with \mbox{10,000} training samples, an improvement of around $40\,\%$ remains in the $50\,\text{GeV}$ case, while for the $20\,\text{GeV}$ images, the SR CNNs only outperform the LR ones for high photon efficiencies ($>95\,\%$).
When training on \mbox{100,000} samples, the performance of the SR and LR CNNs is similar for both energies. 

In an actual experiment, using SR as a pre-processing step for training a photon-identification classifier can indeed be useful.
While large amounts of real single-photon signatures can be easily found in a full simulation (for example from $H\rightarrow\gamma\gamma$ decays), this is typically not the case for fake single-photon candidates. 
Only a tiny fraction of simulated jets leads to signatures which are photon-like, characterized by sharp energy depositions in the ECAL, low hadronic activity close-by and no matched tracks (or a tracker signature compatible with a photon conversion).
Hence, the fraction of simulated jets passing typical photon pre-selection criteria based on shower-shape variables as well as requirements on the photon isolation, i.e., the activity around the photon candidate, is typically very small.
Therefore, the fake single-photon datasets that are available for the classifier trainings are often small.
However, particle-gun simulations of photons and neutral pions, such as those that we used for these studies, can be easily produced in large amounts also with a realistic detector simulation.
If SR networks that are trained on such particle-gun simulations are found to be universal in the sense that they capture the main properties of the electromagnetic showers, they could be used as a pre-processing step for the classifier trainings based on real and fake single photons in the experiment.
We hence propose further studies in this direction.

\FloatBarrier

\section{Conclusions}
\label{sec:conclusions}

We used simulated showers of $20$ and $50\,\text{GeV}$ single photons and neutral-pion decays to two photons in a toy PbWO$_4$ calorimeter to train super-resolution networks based on the ESRGAN architecture.
We treated the energy depositions in the calorimeter crystals as two-dimensional images and created low-resolution images, corresponding to the nominal resolution, and high-resolution counterparts, which correspond to an artificially increased resolution by a factor of four in both dimensions.
We made modifications to the original ESRGAN proposal based on training properties of Wasserstein Generative Adversarial Networks and based on the physics properties of the images.
In particular, we found that a physics-inspired perceptual-loss term improves the training, which we based on the features that convolutional neural networks extracted from the high-resolution images.

We found that the super-resolution networks are able to reproduce distinct features of the high-resolution images, which were not apparent in the low-resolution images by eye, such as the presence of a second energy maximum for the pion decays.
We also found that the networks are able to upsample low-resolution images of photons and pions generally in a convincing way, although the networks are trained on photons and pions together and the label of each image is not explicitly passed to the networks.
We then studied possible applications of the super-resolution images at collider experiments and we found that the reconstruction of the shower width (as an example of a shower-shape variable) and of the position of the shower center are much improved compared to the reconstruction from the low-resolution images.
We also studied whether the super-resolution images could be used as a pre-processing step for training photon-identification classifiers at collider experiments.
When only a low number of samples was available for the classifier training, the training on the super-resolution images outperformed the training on the low-resolution counterparts.
We conclude that the additional physics information that is included in the high-resolution images, and hence also in the generated super-resolution images, helps to extract discriminatory features for the classification.

In general, we conclude that the application of super resolution based on the proposed modified ESRGAN architecture is promising for the analysis of photon signatures at collider experiments.
While the photons' calorimeter signatures are used for several different reconstruction and identification goals, for which typically separate algorithms are trained, the super-resolution is intrinsically multi-purpose and promises to improve several tasks at once.
As one example, we stress the challenge in simulating a sufficient number of fake single-photon candidates from jets at hadron-collider experiments, and the benefits that a pre-processing with a particle-gun-based super-resolution network could bring.
Future studies on super-resolution networks for collider experiments should expand the energy range, use the realistic simulations that are available at the LHC experiments, and study the performance of particle-gun-based super resolution on full collider events.

\subsubsection*{Acknowledgements}

This research was supported by the Deutsche Forschungsgemeinschaft (DFG) under grants 400140256 - GRK~2497 (The physics of the heaviest particles at the LHC, JE and FM) and 686709 - ER~866/1-1 (Heisenberg Programme, JE), by the Studienstiftung des deutschen Volkes (FM), and by the Bundesministerium f\"ur Bildung und Forschung (BMBF) under grant 05H21PECA1 (AvdG and ON).

\clearpage

\bibliographystyle{JHEP-no-scshape}
\bibliography{main}

\end{document}